\documentclass[letterpaper, 10 pt, journal]{IEEEtran} 

\IEEEoverridecommandlockouts                              
\usepackage{amsmath,mathrsfs}
\usepackage{amstext}
\usepackage{color,colortbl}
\usepackage[table]{xcolor}
\usepackage{amsfonts}
\usepackage{amssymb,amsbsy}
\usepackage{graphicx}
\usepackage{epstopdf}
\usepackage{url}
\usepackage[us]{datetime}
\usepackage{cite}
\usepackage[thmmarks, amsmath]{ntheorem}

\allowdisplaybreaks[1]
\graphicspath{{../},{../Bio/}, {../Fig/}, {../Fig05/}, {../Fig06/}, {../Fig07/}}

\definecolor{Gray}{gray}{0.9}
\definecolor{LightCyan}{rgb}{0.88,1,1}
\definecolor{LightMagenta}{rgb}{1,0.88,1}
\definecolor{LightOrange}{rgb}{1,0.99,0.88}
\definecolor{LightGreen}{rgb}{0.9,1,0.8}
\definecolor{DarkOrange}{rgb}{0.98,0.91,0.71}
\definecolor{DarkGreen}{rgb}{0.67,0.88,0.69}

\theorembodyfont{\normalfont}
\theoremseparator{:}
\newtheorem{theorem}{Theorem}
\newtheorem{lemma}{Lemma}
\newtheorem{definition}{Definition}
\newtheorem{remark}{Remark}
\theoremstyle{nonumberplain}
\theoremsymbol{$\square$}

\newcommand{\bfC}[0]{\mathbf{C}}
\newcommand{\bfD}[0]{\mathbf{D}}
\newcommand{\bfH}[0]{\mathbf{H}}

\newcommand{\bfR}[0]{\mathbf{R}}

\newcommand{\bfd}[0]{\boldsymbol{d}}

\newcommand{\bfr}[0]{\boldsymbol{r}}

\newcommand{\cFE}[0]{\mathcal{F}} 
\newcommand{\cFB}[1][]{\mathcal{F}_{#1}^{B}} 

\newcommand{\pvec}[1]{\boldsymbol{r}_{#1}}
\newcommand{\diff}[0]{\textrm{d}}
\newcommand{\unitvec}[1][]{\boldsymbol{e}_{#1}}
\DeclareMathOperator{\vectr}{\overline{vec}}
\DeclareMathOperator{\sign}{sign}

\title{
Representing Lanes as Arc-length-based Parametric Curves to Facilitate Estimation in Vehicle Control
}
\IEEEpubid{{\color{red} This paper is currently under review. This preprint is dedicated for REVIEW ONLY}.}

\author{Wubing~B.~Qin
\thanks{Manuscript revised \currenttime, \today.%
}
\thanks{Wubing B.~Qin is with the Department of Mechanical Engineering, University of Michigan, Ann Arbor, MI 48109, USA. (Email: wubing@umich.edu,).}%
}

\begin{document}
\maketitle

\begin{abstract}
This paper revisits the fundamental mathematics of Taylor series to approximate curves with function representation and arc-length-based parametric representation. Parametric representation is shown to preserve its form in coordinate transformation and parameter shifting. These preservations can significantly facilitate lane estimation in vehicle control since lanes perceived by cameras are typically represented in vehicle body-fixed frames which are translating and rotating. Then we derived the transformation from function representation to arc-length-based parametric representation and its inverse. We applied the transformation to lane estimation in vehicle control problem, and derived the evolution of coefficients for parametric representation that can be used for prediction. We come up with a procedure to simulate the whole process with perception, lane estimation and control for the path-following problem. Simulations are performed to demonstrate the efficacy of the proposed lane estimation algorithm using parametric representation. The results indicate that the proposed technique ensures that vehicle control can achieve reasonably good performance at very low perception updating rate.
\end{abstract}


\section{Introduction}

Last decade witnessed an increased effort in the deployment of automated vehicle (AV) technology because it can enhance passenger safety, improve travel mobility, reduce fuel consumption, and maximize traffic throughput \cite{VanderWerfShladover02, Askari_2016, Li_AAP_2017}. Early deployment dates back to research projects, such as DARPA challenges \cite{Campbell_2010}, PATH program \cite{Rajam_Shlad_2001}, Grand Cooperative Driving Challenges \cite{GCDCIntro2011}. Meanwhile, the automotive industry made achievements in equipping production vehicles with advanced driver assistant systems (ADAS) technology. Currently attentions are attracted to AVs with higher levels of autonomy \cite{SAE_J3016_2016}.

The ultimate objective is to navigate and guide AVs to destinations following driving conventions and rules.
The key components include perception, estimation, planning and control. Perception algorithms collect and process raw sensor (camera, radar, lidar, etc.) data, and convert them to human-readable physical data. Estimation algorithms \cite{Farrell_2017, Wischnewski_2019, Bersani_AEIT_2019} typically apply sensor fusion technique to obtain clean and sound vehicle state estimations based on sensor characteristics. Planning \cite{Karaman_IJRR_2011, Paden_TIV_2016} can be further divided into mission planning (or route planning), behavior planning (or decision making), and motion planning: i) mission planning algorithms select routes to destinations through given road networks based on requirements; ii) behavior planning generates appropriate driving behaviors in real-time according to driving conventions and rules, to guide the interaction with other road-users and infrastructures; iii) motion planning translates the generated behavior primitives into a trajectory based on vehicle states. Control algorithms utilize techniques from control theory \cite{Karl_Richard_Feedback, Khalil_NC, Luenberger_1997, Ioannou_Sun_RAC, Yuri_SMC} that enable vehicles to follow aforementioned trajectories in longitudinal and lateral directions.

\IEEEpubidadjcol

Vehicle control performance highly depends on perception and estimation outcomes, among which lane perception and lane estimation \cite{AHuang_phd, Fakhfakh_JAIHC_2020} are crucial. In longitudinal control, estimated lanes are needed to identify preceding in-path vehicles, and obtain maximum allowable speed based on road curvature at a preview distance. In lateral control, estimated lanes serve as desired paths for controllers to follow. In practice, lanes are typically perceived by cameras in real-time when high-resolution GPS or high-definition map are not installed. Perception algorithms detect lanes using techniques from computer vision or machine learning and then characterize them as polynomial functions \cite{Xu_ECCV_2020, Liu_ICCV_2021_CondLaneNetAT, Wang_ECCV_2020, Tabelini_2021} in vehicle body-fixed frame. However, representing lanes as polynomial functions brings great inconvenience to practical estimation and control algorithms. On the one hand, estimation algorithms usually predict polynomial coefficients based on a nominal model, among which Kalman filter \cite{Klmn_1960, Humpherys_2012, Zarchan_KF_2015, Pei_KF_2019} is the most famous. However, it is infeasible to mathematically derive such a model that characterizes the evolution of coefficients for polynomial functions. This is because: i) vehicle body-fixed frame is translating and rotating as the vehicle moves, and ii) polynomial function representation does not preserve the form in coordinate transformation.
On the other hand, iterative methods need to be applied when control algorithms extract attributes (curvature, heading, etc.) from polynomial function representation of lanes at a preview distance.

This paper attempts to resolve this problem by characterizing lanes with arc-length-based parametric representation. To ensure compatibility with current platforms, we still assume that perception algorithms provide polynomial function representation of lanes in vehicle body-fixed frame.
The major contributions are as follows. Firstly, we mathematically derived a transformation and its inverse that reveal the relationship between polynomial function representation and arc-length-based parametric representation. Secondly, it is shown that such parametric representation preserves the form in coordinate transformation. Therefore, we are able to derive a mathematical model to characterize the evolution of coefficients that can be used for prediction during locomotion. Moreover, to simulate the whole process of lane perception, lane estimation and control, we set up a novel simulation framework that includes: i) usage of curvature as a function of arc length to represent lanes based on differential geometry, ii) derivation of coefficients for polynomial function representation to simulate perception in vehicle body-fixed frame, and iii) transformation from absolute dynamics in earth-fixed frame to relative dynamics observable in camera-based control.

This paper is organized follows. In Section~\ref{sec:prelim}, we start with preliminaries on coordinate transformation, vectorization operator and fundamental theory on curves. In Section~\ref{sec:curv_repr}, we provide theoretical results on curve representations, and derive transformations between polynomial function representation and arc-length-based parametric representation.
Section~\ref{sec:veh_ctrl_setup} investigates camera-based vehicle control problem, and applies theoretical results in Section~\ref{sec:curv_repr} to obtain an intrinsic linear model for lane estimation. Also, the dynamics describing absolute position and orientation in earth-fixed frame are transformed into those describing relative position and orientation observed in the camera field-of-view (FOV). A controller is introduced from \cite{Wubing_LC_TIV_2022} to demonstrate that prediction based on the proposed lane estimation approach is adequate for control algorithms. In Section~\ref{sec:res}, we first show how we set up simulation framework, and then conduct experiments to demonstrate the efficacy of the simulation framework and lane estimation approach. In Section~\ref{sec:conclusion}, conclusions are drawn and future research directions are pointed out.

\section{Preliminaries\label{sec:prelim}}
In this section, preliminaries are provided. In Section~\ref{sec:coord_transf_gen}, we provide the coordinate transformation applied extensively in later chapters. In Section~\ref{sec:vec_op}, the definition and theorem on vector operator of matrices are introduced. In Section~\ref{sec:curv_repr_def}, polynomial function representation and arc-length-based parametric representation are discussed to approximate curves based on Taylor series.

\subsection{Coordinate Transformation \label{sec:coord_transf_gen}}

\begin{figure}
  \centering
  \includegraphics[scale=0.5]{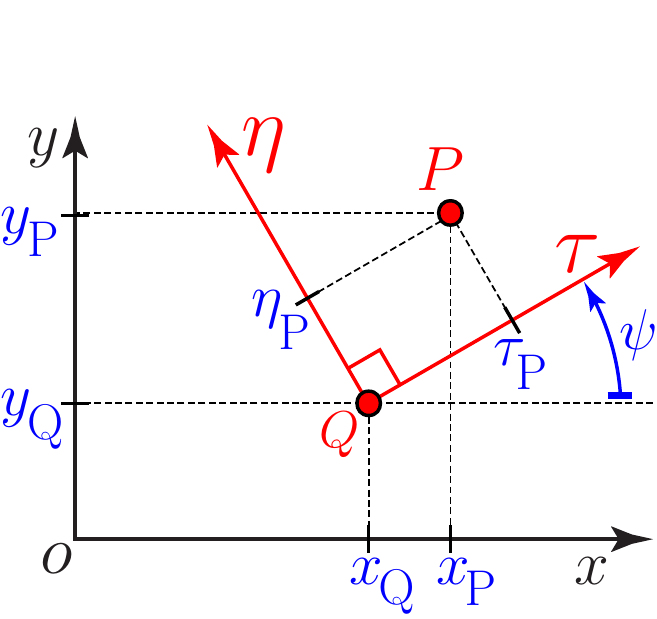}\\
  \caption{Coordinate transformation.\label{fig:gen_coord_transf}}
\end{figure}

Given two coordinate systems ($O$-$xyz$ and $Q$-$\tau\eta z$) as shown in Fig.~\ref{fig:gen_coord_transf}, where the $Q$-$\tau\eta z$ frame is obtained by translating the $O$-$xyz$ frame with vector $\pvec{OQ}$, and then rotating with angle $\psi$, we recall the following theorem.

\begin{theorem}[Change of Coordinates]\label{thm:gen_coord_transf}
  Suppose the coordinates of $Q$ expressed in $O$-$xyz$ frame are
  \begin{align}
    \boldsymbol{d} &=
    \begin{bmatrix}
      x_{\rm Q} & y_{\rm Q}
    \end{bmatrix}^{\top}\,,
  \end{align}
  and the coordinates of an arbitrary point $P$ are
  \begin{align}
    \boldsymbol{r} &=
    \begin{bmatrix}
      x_{\rm P} & y_{\rm P}
    \end{bmatrix}^{\top}\,,&
    \hat{\boldsymbol{r}} &=
    \begin{bmatrix}
      \tau_{\rm P} & \eta_{\rm P}
    \end{bmatrix}^{\top}\,,
  \end{align}
  expressed in the $O$-$xyz$ and $Q$-$\tau\eta z$ frame, respectively, then one can change coordinates from $Q$-$\tau\eta z$ frame to $O$-$xyz$ frame by
  \begin{align}
    \boldsymbol{r} &= \bfR\, \hat{\boldsymbol{r}}+\boldsymbol{d}\,,
  \end{align}
  where
  \begin{align}
    \bfR &=
    \begin{bmatrix}
      \cos\psi  & -\sin\psi  \\ \sin\psi  &\cos\psi
    \end{bmatrix}
  \end{align}
  is the rotation matrix.
\end{theorem}

\subsection{Vectorization of Matrices \label{sec:vec_op}}
Vectorization of a matrix represents the operation that cuts the matrix into columns and stacks them sequentially as a column vector.
One can refer to \cite{Laub2005} for more details. Here we recall the following definition and theorem.

\begin{definition}\label{def:vec_op}
Let $h_{i}\in \mathbb{R}^{n}$ denote the $i$-th column of matrix $\bfH\in \mathbb{R}^{n\times m}$, i.e., $\bfH=[h_{1}\quad h_{2}\quad\ldots\quad
h_{m}]$. The vector operator is defined as
\begin{equation}\label{eqn:vec_def}
  \vectr(\bfH)=
  \left[
    \begin{array}{cccc}
      h_{1}^{\rm T} & h_{2}^{\rm T} &\ldots & h_{m}^{\rm T}
    \end{array}
  \right]^{\rm T}\in
\mathbb{R}^{mn}\,.
\end{equation}
\end{definition}

\begin{theorem}\label{thm:vec_kron_relation}
For any three matrices $\mathbf{A}$, $\mathbf{B}$ and $\bfC$ where the matrix product
$\mathbf{ABC}$ is defined, we have
\begin{equation}
  \vectr(\mathbf{A}\mathbf{B}\bfC)=(\bfC^{\rm T}\otimes \mathbf{A})\vectr(\mathbf{B})\,,
\end{equation}
where $\otimes$ denotes Kronecker product.
\end{theorem}

\subsection{Curves in 2-D Space \label{sec:curv_repr_def}}
According to Taylor series, curves can be approximated with polynomials.
In practice, lanes are typically represented as 2-D curves in vehicle body-fixed frame.
Thus, in this part we discuss Taylor approximation for 2-D curves in $xy$-plane. The shorthand notation
\begin{align}
  \dfrac{\diff^{0} y}{\diff x^{0}}=y(x)\,,
\end{align}
is introduced, which will be kept throughout this paper.

\begin{definition}
  Given a curve $\mathcal{C}$ in $xy$-plane and a point $P$ on the curve whose coordinates are $(x_{0}, y(x_{0}))$, its $N$-th order polynomial function representation about point $P$ is its Taylor approximation at point $P$ until the $N$-th order, i.e.,
  \begin{equation}
  \begin{split}\label{eqn:def_func_repr_algb_form}
    \mathcal{C}'&=\{(x,\, y)\,|\,
    y(x;x_{0})=\varphi_{0}+\varphi_{1}\, (x-x_{0}) +\cdots \\
    &+\varphi_{N}\, (x-x_{0})^{N}\,,\,x\in\mathbb{R}\}\,,
  \end{split}
  \end{equation}
  where the coefficients are
  \begin{align}\label{eqn:func_repr_coeff_def}
    \varphi_{n} &= \dfrac{1}{n!}\dfrac{\diff^{n} y}{\diff x^{n}}\Big|_{x=x_{0}}\,,
  \end{align}
  for $n=0, 1, \ldots, N$.
\end{definition}

\begin{definition}
  Given a curve $\mathcal{C}$ in $xy$-plane and a point $P$ on the curve, the $N$-th order parametric representation using arc length
  \begin{align}\label{eqn:arc_length_def}
    s &= \int_{x_{0}}^{x}\sqrt{1+\Big(\dfrac{\diff y}{\diff x}\Big)^{2}}\, \diff x\,,
  \end{align}
  about point $P$ is its Taylor approximation with respect to arc length parameter $s$ at point $P$ until the $N$-th order, i.e.,
  \begin{equation}\label{eqn:def_param_repr_algb_form}
    \begin{split}
     \mathcal{C}'&=\{(x,\, y)\,|\,
      x(s; s_{0})=\bar{\phi}_{0}+\bar{\phi}_{1} \,(s-s_{0})+\cdots\\
      & +\bar{\phi}_{N} \,(s-s_{0})^{N},\;
      y(s; s_{0})=\hat{\phi}_{0}+\hat{\phi}_{1}\, (s-s_{0})\\
      & +\cdots +\hat{\phi}_{N}\, (s-s_{0})^{N},\,s\in\mathbb{R}\}\,
    \end{split}
  \end{equation}
  where $s_{0}$ is the corresponding arc length of point $P$, and the coefficients are
  \begin{align}\label{eqn:param_repr_coeff_def}
    \bar{\phi}_{n}&= \dfrac{1}{n!}\dfrac{\diff^{n} x}{\diff s^{n}}\Big|_{s=s_{0}}\,,&
    \hat{\phi}_{n}&= \dfrac{1}{n!}\dfrac{\diff^{n} y}{\diff s^{n}}\Big|_{s=s_{0}}\,,
  \end{align}
  for $n=0, 1, \ldots, N$.
\end{definition}

We remark that in the  remainder of the paper polynomial function representation \eqref{eqn:def_func_repr_algb_form} and arc-length-based parametric representation \eqref{eqn:def_param_repr_algb_form} will be referred to as function representation and parametric representation, respectively. To simplify the notation, we rewrite \eqref{eqn:def_func_repr_algb_form} into the matrix form, that is,
\begin{align}\label{eqn:def_func_repr_matx_form}
  \mathcal{C}'&=\{(x,\, y)\,|\,y(x; x_{0})=\boldsymbol{\varphi}(x_{0})\,\boldsymbol{p}_{N}(x-x_{0})\,,\,x\in\mathbb{R}\}\,,
\end{align}
where
\begin{equation}
\begin{split}\label{eqn:func_repr_coeff_vec_def}
    \boldsymbol{\varphi} & =
    \begin{bmatrix}
      \varphi_{0} &\varphi_{1} & \cdots & \varphi_{N}
    \end{bmatrix}\,,\\
    \boldsymbol{p}_{N}(x) & =
    \begin{bmatrix}
      1 & x & \cdots & x^{N}
    \end{bmatrix}^\top \,.
\end{split}
\end{equation}
Similarly, the matrix form of \eqref{eqn:def_param_repr_algb_form} is
\begin{align}\label{eqn:def_param_repr_matx_form}
  \mathcal{C}'&=\{\boldsymbol{r}\,|\,
  \boldsymbol{r}(s; s_{0})=\boldsymbol{\phi}(s_{0})\,\boldsymbol{p}_{N}(s-s_{0})\,,\,s\in\mathbb{R}\}\,,
\end{align}
where
\begin{equation}\label{eqn:param_repr_coeff_vec_def}
\begin{split}
    \boldsymbol{r}(s; s_{0}) & =
    \begin{bmatrix}
      x(s; s_{0}) \\ y(s; s_{0})
    \end{bmatrix}\,,\enskip
    \boldsymbol{\phi}  =
    \begin{bmatrix}
      \bar{\phi}_{0} & \bar{\phi}_{1} & \cdots & \bar{\phi}_{N} \\
      \hat{\phi}_{0} & \hat{\phi}_{1} & \cdots & \hat{\phi}_{N}
    \end{bmatrix}\,,
\end{split}
\end{equation}
and $\boldsymbol{p}_{N}(\cdot)$ uses the same definition in \eqref{eqn:func_repr_coeff_vec_def}. Note that the dependency of coefficients ($\boldsymbol{\varphi}$ and $\boldsymbol{\phi}$) on the coordinates of point P ($x_{0}$ and $s_{0}$) is highlighted in the matrix form.

\begin{remark}\label{remark:curv_repr_gen}
  In 2-D space, a curve is uniquely determined if the curvature $\kappa(s)$ is given for arbitrary arc length position $s$.
\end{remark}
\begin{IEEEproof}
  For a general point $P$ on the curve, we assume that its coordinates, slope angle, curvature and arc length are $(x, y)$, $\alpha$, $\kappa$ and $s$, respectively.
  Based on differential geometry, their relationship is described by
  \begin{equation}\label{eqn:EOM_ref_path_ds}
    \begin{split}
         x' &=\cos\alpha\, ,
         \\
        y' &=\sin\alpha\, ,
        \\
        \alpha' &=\kappa\, ,
    \end{split}
  \end{equation}
  where prime denotes differentiation with respect to $s$. Solving \eqref{eqn:EOM_ref_path_ds} with initial conditions, one can obtain the tuple $(x, y,\alpha)$ as functions of $s$ that characterizes the curve.
\end{IEEEproof}

\section{Representations of Curves \label{sec:curv_repr}}
This part presents theoretical results on the representations of curves. Section~\ref{sec:property_param_repr} shows that parametric representations have property of preserving the form in parameter shifting and coordinate transformation. Section~\ref{sec:transform_repr} derives the changes of coefficients that reveal the relationship between function representation and parametric representation. In Section~\ref{sec:coeff_repr_curves}, we derive the coefficients for both representations for the curve \eqref{eqn:EOM_ref_path_ds} with known curvature at arbitrary arc length position.

\subsection{Properties of Parametric Representation \label{sec:property_param_repr}}
\begin{theorem}[Conformal Representation in Shifting]\label{thm:curve_param_repr_shift}
   The parametric curve \eqref{eqn:def_param_repr_matx_form} preserves its form when the Taylor expansion is shifted to parameter location $s_{1}=s_{0}+\tilde{s}$. That is, $\boldsymbol{r}(s; s_{0})$ can be rewritten into
  \begin{align}
    \boldsymbol{r}(s; s_{1}) &= \boldsymbol{\phi}(s_{1})\,\boldsymbol{p}_{N}(s-s_{1})\,,
  \end{align}
  where the change of coefficients is
  \begin{align}
    \boldsymbol{\phi}^{\top}(s_{1}) &= \boldsymbol{T}(\tilde{s})\,\boldsymbol{\phi}^{\top}(s_{0})\,,
  \end{align}
  and
  \begin{align}\label{eqn:T_expansion}
    \boldsymbol{T}(\tilde{s})&=
    \begin{bmatrix}
    C_{0}^{0} &  C_{1}^{0}\tilde{s} & C_{2}^{0}\tilde{s}^{2} & \cdots & C_{N}^{0}\tilde{s}^{N}\\
     &  C_{1}^{1} & C_{2}^{1}\tilde{s} &\cdots & C_{N}^{1}\tilde{s}^{N-1}\\
     &  & C_{2}^{2} &  \ddots & \vdots \\
     &  &  & \ddots  & C_{N}^{N-1}\tilde{s}& \\
     &  &  &  & C_{N}^{N}
    \end{bmatrix}.
  \end{align}
  Here, $C_{N}^{k}$ represents the binomial coefficients and $C_{0}^{0}=1$.
\end{theorem}
\begin{IEEEproof}
  Substituting ${s_{0}=s_{1}-\tilde{s}}$ into \eqref{eqn:def_param_repr_matx_form}, and then utilizing binomial theorem, one can obtain
\begin{align}
  \boldsymbol{r}&(s; s_{0})= \boldsymbol{\phi}(s_{0})\,\boldsymbol{p}_{N}(s-s_{0}) \nonumber\\
  &=
  \begin{bmatrix}
    \sum\limits_{n=0}^{N} \bar{\phi}_{n} (s-s_{0})^{n}\\
    \sum\limits_{n=0}^{N} \hat{\phi}_{n} (s-s_{0})^{n}
  \end{bmatrix}
  =
  \begin{bmatrix}
    \sum\limits_{n=0}^{N} \bar{\phi}_{n} (s-s_{1}+\tilde{s})^{n}\\
    \sum\limits_{n=0}^{N} \hat{\phi}_{n} (s-s_{1}+\tilde{s})^{n}
  \end{bmatrix} \nonumber\\
  &=
  \begin{bmatrix}
    \sum\limits_{n=0}^{N} \bar{\phi}_{n} \sum\limits_{m=0}^{n}C_{n}^{m}\tilde{s}^{n-m}(s-s_{1})^{m}\\
    \sum\limits_{n=0}^{N} \hat{\phi}_{n} \sum\limits_{m=0}^{n} C_{n}^{m}\tilde{s}^{n-m}(s-s_{1})^{m}
  \end{bmatrix}\\
  &=
  \begin{bmatrix}
    \sum\limits_{m=0}^{N} \Big(\sum\limits_{n=m}^{N} C_{n}^{m}\tilde{s}^{n-m}\bar{\phi}_{n}\Big)(s-s_{1})^{m}\\
    \sum\limits_{m=0}^{N} \Big(\sum\limits_{n=m}^{N} C_{n}^{m}\tilde{s}^{n-m}\hat{\phi}_{n}\Big)(s-s_{1})^{m}
  \end{bmatrix}\,,\nonumber
\end{align}
which is the $N$-th order parametric curve about $s_{1}$. Comparing the coefficients, we obtain the change of coefficients.
\end{IEEEproof}

\begin{theorem}[Conformal Representation in Transformation] \label{thm:curve_param_repr_conformal}
  The parametric curve \eqref{eqn:def_param_repr_matx_form} preserves its form in coordinate transformation. Suppose we are given two coordinate systems (frame $\cFE_{1}$ and $\cFE_{2}$), and the change of coordinates from $\cFE_{2}$ to $\cFE_{1}$ is
  \begin{align}
    \boldsymbol{r} &= \bfR\,\hat{\boldsymbol{r}} + \bfd\,,
  \end{align}
  where $\boldsymbol{r}$ and $\hat{\boldsymbol{r}}$ are the coordinates of an arbitrary point expressed in frame $\cFE_{1}$ and $\cFE_{2}$, respectively, $\bfR$ is the rotation matrix, and $\bfd$ is the origin of frame $\cFE_{2}$ expressed in frame $\cFE_{1}$.
  Then the parametric curve \eqref{eqn:def_param_repr_matx_form} expressed in frame $\cFE_{1}$ possesses the same form when expressed in frame $\cFE_{2}$, that is,
  \begin{align}
    \hat{\boldsymbol{r}}(s; s_{0}) &=\widehat{\boldsymbol{\phi}}(s_{0})\,\boldsymbol{p}_{N}(s-s_{0})\,,
  \end{align}
  where the change of coefficients is
  \begin{align}
    \widehat{\boldsymbol{\phi}}(s_{0})& = \bfR^{\top}\big(\boldsymbol{\phi}(s_{0})-\bfD \big)\,,
  \end{align}
  and
  \begin{align}
    \bfD& = \begin{bmatrix}
      \bfd & 0 & \cdots & 0
    \end{bmatrix}\in \mathbb{R}^{2\times (N+1)}\,.
  \end{align}
\end{theorem}
\begin{IEEEproof}
  Based on change of coordinates, the curve \eqref{eqn:def_param_repr_matx_form} can be expressed in $\cFE_{2}$ as
\begin{equation}
  \begin{split}
    \hat{\boldsymbol{r}}(s; s_{0}) &= \bfR^{\top}\boldsymbol{r}(s; s_{0})- \bfR^{\top}\bfd\,,\\
    &=\bfR^{\top}\boldsymbol{\phi}(s_{0})\,\boldsymbol{p}_{N}(s-s_{0}) - \bfR^{\top}\bfd\,,
  \end{split}
\end{equation}
By noticing that
\begin{align}
  \bfd & = \bfD\,\boldsymbol{p}_{N}(s-s_{0})\,,
\end{align}
one can obtain the change of coefficients given in the theorem.

\end{IEEEproof}

\begin{remark}\label{remark:func_repr_preserve_shift}
  Function representation preserves its form in parameter shifting, but not in coordinate transformation.
\end{remark}

\subsection{Transformations between Function Representation and Parametric Representation \label{sec:transform_repr}}

In this part, we derive the transformations between function representation and parametric representation.
Results are only provided until the fifth-order, which is considered adequate in practice.
The following assumptions are made: i) Taylor series are expanded about the intersection between the curve and the $y$-axis (also referred to as $y$-intercept); ii) this $y$-intercept marks the starting point of arc length, i.e., $s=0$; and iii) the positive direction of arc length parameter $s$ corresponds to the positive direction of the $x$-axis. We obtain the following transformations based on those assumptions.

\begin{theorem}\label{thm:curve_func_2_param_repr}
  Given a curve with function representation \eqref{eqn:def_func_repr_matx_form}, its parametric representation \eqref{eqn:def_param_repr_matx_form} is uniquely determined in the same coordinate system. In other words, there exists a unique map $\boldsymbol{\phi}=\boldsymbol{f}(\boldsymbol{\varphi})$. The coefficients until the fifth order are
  \begin{equation}
    \begin{split}
    \bar{\phi}_{0} &= 0\,, \qquad\qquad\qquad\qquad\quad\enskip\;
    \hat{\phi}_{0} = \varphi_{0}\,,\\
    \bar{\phi}_{1} &= \dfrac{1}{\lambda}\,, \qquad\qquad\qquad\qquad\quad\enskip
    \hat{\phi}_{1} = \dfrac{\varphi_{1}}{\lambda}\,,\\
    \bar{\phi}_{2} &= -\dfrac{\varphi_{1} \varphi_{2}}{\lambda^{4}}\,, \qquad\qquad\qquad\quad
    \hat{\phi}_{2} =\dfrac{\varphi_{2}}{\lambda^{4}}\,,\\
    \bar{\phi}_{3} &= \dfrac{2 \varphi_{2}^{2}-\varphi_{1} \varphi_{3}}{\lambda^{5}}-\dfrac{8 \varphi_{2}^{2}}{3\lambda^{7}}\,,\qquad
    \hat{\phi}_{3} =\dfrac{\varphi_{3}}{\lambda^{5}}- \dfrac{8\varphi_{1} \varphi_{2}^{2}}{3\lambda^{7}}\,, \\
    \bar{\phi}_{4} &= \dfrac{5 \varphi_{2} \varphi_{3}-\varphi_{1} \varphi_{4}}{\lambda^{6}}
    -\dfrac{10 \varphi_{1} \varphi_{2}^{3}+13 \varphi_{2} \varphi_{3}}{2\lambda^{8}}\\
     &+\dfrac{28\varphi_{1} \varphi_{2}^{3}}{3\lambda^{10}}\,,\\
    \hat{\phi}_{4} &=\dfrac{\varphi_{4}}{\lambda^{6}}
    +\dfrac{16 \varphi_{2}^{3}-13 \varphi_{1} \varphi_{2}\varphi_{3}}{2\lambda^{8}}
    - \dfrac{28\varphi_{2}^{3}}{3\lambda^{10}} \,,\\
    \bar{\phi}_{5} &= \dfrac{3 \varphi_{3}^{2}+6 \varphi_{2} \varphi_{4}-\varphi_{1} \varphi_{5}}{\lambda^{7}}\\
    &-\dfrac{39 \varphi_{3}^{2}+210 \varphi_{1}\varphi_{2}^{2} \varphi_{3}+76 \varphi_{2} \varphi_{4}-140 \varphi_{2}^{4}}{10\lambda^{9}}\\
    &+\dfrac{188 \varphi_{1} \varphi_{2}^{2}\varphi_{3}-248 \varphi_{2}^{4}}{5\lambda^{11}}
    + \dfrac{112\varphi_{2}^{4}}{3\lambda^{13}}\,,\\
    \hat{\phi}_{5} &=\dfrac{\varphi_{5}}{\lambda^{7}}
    +\dfrac{326 \varphi_{2}^{2} \varphi_{3}-76 \varphi_{1}\varphi_{2} \varphi_{4}-39 \varphi_{1}\varphi_{3}^{2}}{10\lambda^{9}}\\
    &-\dfrac{128 \varphi_{1}\varphi_{2}^{4}+188 \varphi_{2}^{2} \varphi_{3}}{5\lambda^{11}}+ \dfrac{112\varphi_{1}\varphi_{2}^{4} }{3\lambda^{13}}\,,
    \end{split}
  \end{equation}
  where
  \begin{align}
    \lambda &= \sqrt{1+\varphi_{1}^{2}}\,.
  \end{align}
\end{theorem}

\begin{IEEEproof}
  Based on the assumptions, we have
\begin{align}\label{eqn:fun_repr_eval_0}
   x (0)&=0\,, &
   y (0)&=\varphi_{0}\,,&
   x '(s) &>0\,.
\end{align}
where $'$ denotes differentiation with respect to $s$, implying $\bar{\phi}_{0}$ and $\hat{\phi}_{0}$ given in the theorem.
Notice that \eqref{eqn:arc_length_def} yields
\begin{align}\label{eqn:ds_comp_equality}
  \diff s^{2} &=\diff  x ^{2}+\diff  y ^{2} &\Longrightarrow \quad
  ( x ')^{2}+( y ')^{2} &=1\,,
\end{align}
and the derivative of \eqref{eqn:def_func_repr_algb_form} with respect to $s$ yields
\begin{align}\label{eqn:fun_repr_deriv_1}
   y ' &= \varphi_{1}  x '+\cdots +n\varphi_{n}  x ^{n-1} x '\,.
\end{align}
Evaluating (\ref{eqn:ds_comp_equality}, \ref{eqn:fun_repr_deriv_1}) at $s=0$ and utilizing \eqref{eqn:fun_repr_eval_0}, we obtain
\begin{align}\label{eqn:fun_repr_deriv1_eval_0}
   x '(0) &=\dfrac{1}{\sqrt{1+\varphi_{1}^{2}}}\,,&
   y '(0) &=\dfrac{\varphi_{1}}{\sqrt{1+\varphi_{1}^{2}}}\,,
\end{align}
which implies $\bar{\phi}_{1}$ and $\hat{\phi}_{1}$ given in the theorem. Then taking the derivatives of (\ref{eqn:ds_comp_equality}, \ref{eqn:fun_repr_deriv_1}) with respect to $s$ yields
\begin{equation}\label{eqn:fun_repr_deriv2}
  \begin{split}
    & x ' x ''+ y ' y '' =0\,,\\
    & y '' = (\varphi_{1} +\cdots +n\varphi_{n}  x ^{n-1}) x ''\\
    &\quad +\big(2\varphi_{2} +\cdots +n(n-1)\varphi_{n}  x ^{n-2}\big)( x ')^{2}\,.
  \end{split}
\end{equation}
Evaluating (\ref{eqn:fun_repr_deriv2}) at $s=0$ and utilizing (\ref{eqn:fun_repr_eval_0}, \ref{eqn:fun_repr_deriv1_eval_0}), we obtain
\begin{align}\label{eqn:fun_repr_deriv2_eval_0}
   x ''(0) &=-\dfrac{2\varphi_{1}\varphi_{2}}{(1+\varphi_{1}^{2})^{\frac{3}{2}}}\,,&
   y ''(0) &=\dfrac{2\varphi_{2}}{(1+\varphi_{1}^{2})^{\frac{3}{2}}}\,,
\end{align}
which implies $\bar{\phi}_{2}$ and $\hat{\phi}_{2}$ given in the theorem.
Similarly, evaluating the derivative of (\ref{eqn:fun_repr_deriv2}) at $s=0$ and then utilizing (\ref{eqn:fun_repr_eval_0}, \ref{eqn:fun_repr_deriv1_eval_0}, \ref{eqn:fun_repr_deriv2_eval_0}), one can obtain an algebraic linear equation about $ x '''(0)$ and $ y '''(0)$ and thus obtain $\bar{\phi}_{3}$ and $\hat{\phi}_{3}$ . Following this procedure, we can derive all the derivatives and obtain the coefficients given in the theorem.

\end{IEEEproof}

\begin{theorem}\label{thm:curve_param_2_func_repr}
  Given a curve with parametric representation \eqref{eqn:def_param_repr_matx_form}, its function representation \eqref{eqn:def_func_repr_matx_form} is uniquely determined in the same coordinate system. In other words, there exists a unique map $\boldsymbol{\varphi}=\boldsymbol{f}^{-1}(\boldsymbol{\phi})$.  The coefficients until the fifth-order are
  \begin{equation}
    \begin{split}
      \varphi_{0} &= \hat{\phi}_{0}\,,\qquad\,
      \varphi_{2} = \dfrac{\hat{\phi}_{2} \bar{\phi}_{1}-\hat{\phi}_{1} \bar{\phi}_{2}}{\bar{\phi}_{1}^3}\,,\\
      \varphi_{1} &= \dfrac{\hat{\phi}_{1}}{\bar{\phi}_{1}}\,,\qquad
      \varphi_{3} = \dfrac{\hat{\phi}_{3}}{\bar{\phi}_{1}^3}
      -\dfrac{\hat{\phi}_{1} \bar{\phi}_{3}+2 \hat{\phi}_{2} \bar{\phi}_{2}}{\bar{\phi}_{1}^4}
      +\dfrac{2 \hat{\phi }_{1} \bar{\phi}_{2}^2}{\bar{\phi}_{1}^5}\,,\\
      \varphi_{4} &= \dfrac{\hat{\phi}_{4}}{\bar{\phi}_{1}^4}
      -\dfrac{\hat{\phi}_{1} \bar{\phi}_{4}+2 \hat{\phi}_{2} \bar{\phi}_{3}+3 \hat{\phi}_{3} \bar{\phi}_{2}}{\bar{\phi}_{1}^5}\\
      &+\dfrac{5 \bar{\phi}_{2}(\hat{\phi}_{1} \bar{\phi}_{3} + \hat{\phi}_{2} \bar{\phi}_{2})}{\bar{\phi}_{1}^6}
      -\dfrac{5 \hat{\phi}_{1} \bar{\phi}_{2}^3}{\bar{\phi}_{1}^7}\,,\\
      \varphi_{5} &=\dfrac{\hat{\phi}_{5}}{\bar{\phi}_{1}^5}
      -\dfrac{\hat{\phi }_{1} \bar{\phi}_{5}+2 \hat{\phi}_{2} \bar{\phi}_{4}+3 \hat{\phi }_{3} \bar{\phi}_{3}+4 \hat{\phi}_{4} \bar{\phi}_{2}}{\bar{\phi }_{1}^6}\\
      &+\dfrac{3 \hat{\phi}_{1} (\bar{\phi}_{3}^2+2 \bar{\phi}_{2} \bar{\phi}_{4})+3\bar{\phi}_{2}(3 \hat{\phi}_{3} \bar{\phi}_{2}+4 \hat{\phi}_{2} \bar{\phi}_{3})}{\bar{\phi}_{1}^7}\\
      &-\dfrac{7 \bar{\phi}_{2}^{2}(3 \hat{\phi}_{1} \bar{\phi}_{3} +2 \hat{\phi}_{2} \bar{\phi}_{2})}{\bar{\phi}_{1}^8}
      +\dfrac{14 \hat{\phi}_{1} \bar{\phi}_{2}^4}{\bar{\phi}_{1}^9}\,.
    \end{split}
  \end{equation}
\end{theorem}

\begin{IEEEproof}
  Given the parametric representation \eqref{eqn:def_param_repr_algb_form} of the curve, one can obtain the derivatives ($x'$, $y'$, $x''$, $y''$, ...) with respect to $s$ until the required order. Also, notice that
\begin{equation}\label{eqn:proof_dyx}
    \dfrac{\diff y}{\diff x} =\dfrac{y'}{x'}\,,
\end{equation}
and
\begin{equation}\label{eqn:proof_dyx_recurv}
    \dfrac{\diff^{n+1} y}{\diff x^{n+1}} =\dfrac{\diff}{\diff x}\left(\dfrac{\diff^{n} y}{\diff x^{n}}\right)
    =\dfrac{\left(\frac{\diff^{n} y}{\diff x^{n}}\right)'}{ x'}\,, \quad n=1,\, 2,\, 3,\, \ldots
\end{equation}
Calculating the derivatives in (\ref{eqn:proof_dyx}, \ref{eqn:proof_dyx_recurv}) recursively and utilizing \eqref{eqn:func_repr_coeff_def}, one can obtain the coefficients given in the theorem by noticing that $x_0=0$ implies $s=0$.

\end{IEEEproof}

\begin{remark}
The transformed representation is not necessarily equal to the original representation due to truncation errors in Taylor series, but can provide a very good approximation. This is because their derivatives are equal up until the specified order at the point where Taylor series are expanded.
\end{remark}

\subsection{Representations of a Given Curve \label{sec:coeff_repr_curves}}
It is useful to derive the coefficients of parametric representation and function representation for curves described by \eqref{eqn:EOM_ref_path_ds} with given $\kappa(s)$. On the one hand, these coefficients can be used to validate the transformations derived in Section~\ref{sec:transform_repr}. On the other hand, they can be used to simulate perception outcomes for camera-based control. In the following, we assume: i) the tuple $(x(s), y(s),\alpha(s))$ are obtained with the given $\kappa(s)$ according to \eqref{eqn:EOM_ref_path_ds}; and ii) the representations are expanded about point $P$ whose arc length position is $s_{0}$. Other attributes at point P are
\begin{equation}\label{eqn:gen_curv_P_loc}
\begin{split}
    &x(s_{0}) =x_{0}\,, \enskip
    y(s_{0}) =y_{0}\,, \enskip
    \alpha(s_{0}) =\alpha_{0}\,, \enskip
    \kappa(s_{0}) =\kappa_{0}\,,\\
    &\dfrac{\diff \kappa}{\diff s}(s_{0})=\kappa'_{0}\,,\quad
    \dfrac{\diff^{2} \kappa}{\diff s^{2}}(s_{0})=\kappa''_{0}\,,\quad
    \dfrac{\diff^{3} \kappa}{\diff s^{3}}(s_{0})=\kappa'''_{0}\,.
\end{split}
\end{equation}

\begin{theorem}\label{thm:gen_curve_param_repr}
  Given a curve in $xy$-plane with known ($x(s)$, $y(s)$, $\alpha(s)$, $\kappa(s)$), its parametric representation about point $P$ is unique. The coefficients until fifth-order are
    \begin{equation} 
      \begin{split}
        \bar{\phi}_{0}&=x_{0}\,,\qquad \qquad \qquad\qquad
        \hat{\phi}_{0}=y_{0}\,,\\
        \bar{\phi}_{1}&=\cos\alpha_{0}\,,\qquad\hspace{45pt}
        \hat{\phi}_{1}=\sin\alpha_{0}\,,\\
        \bar{\phi}_{2}&=-\tfrac{1}{2}\kappa_{0}\sin\alpha_{0}\,,\quad\hspace{31pt}
        \hat{\phi}_{2}=\tfrac{1}{2}\kappa_{0}\cos\alpha_{0}\,,\\
        \bar{\phi}_{3}&=-\tfrac{1}{6}\kappa_{0}^{2}\cos\alpha_{0}-\tfrac{1}{6}\kappa_{0}'\sin\alpha_{0}\,,\\
        \hat{\phi}_{3}&=-\tfrac{1}{6}\kappa_{0}^{2}\sin\alpha_{0}+\tfrac{1}{6}\kappa_{0}'\cos\alpha_{0}\,,\\
        \bar{\phi}_{4}&=-\tfrac{1}{24}(\kappa_{0}''-\kappa_{0}^{3})\sin\alpha_{0}-\tfrac{1}{8}\kappa_{0}\kappa_{0}'\cos\alpha_{0}\,,\\
        \hat{\phi}_{4}&=\tfrac{1}{24}(\kappa_{0}''-\kappa_{0}^{3})\cos\alpha_{0}-\tfrac{1}{8}\kappa_{0}\kappa_{0}'\sin\alpha_{0}\,,\\
        \bar{\phi}_{5}&=\tfrac{1}{120}\big(\kappa_{0}^{4}-3(\kappa_{0}')^{2}-4\kappa_{0}\kappa_{0}''\big)\cos\alpha_{0}\\
        &\quad+\tfrac{1}{120}(6\kappa_{0}^{2}\kappa_{0}'-\kappa_{0}''')\sin\alpha_{0}\,,\\
        \hat{\phi}_{5}&=\tfrac{1}{120}\big(\kappa_{0}^{4}-3(\kappa_{0}')^{2}-4\kappa_{0}\kappa_{0}''\big)\sin\alpha_{0}\\
        &\quad-\tfrac{1}{120}(6\kappa_{0}^{2}\kappa_{0}'-\kappa_{0}''')\cos\alpha_{0}\,.
      \end{split}
    \end{equation}
\end{theorem}
\begin{IEEEproof}
  Given an arbitrary point $(x, y)$ on the curve (\ref{eqn:EOM_ref_path_ds}), the derivatives of the curve at that point are
\begin{equation}\label{eqn:xy_derivs}
  \begin{split}
    x' &= \cos\alpha\,,\qquad\qquad\qquad\quad\enskip
    y' =\sin\alpha\,,\\
    x''&=-\kappa\sin\alpha\,,\qquad\qquad\qquad
    y''=\kappa\cos\alpha\,,\\
    x'''&=-\kappa^{2}\cos\alpha-\kappa'\sin\alpha\,,\\
    x^{(4)}&=-(\kappa''-\kappa^{3})\sin\alpha-3\kappa\kappa'\cos\alpha\,,\\
    x^{(5)}&=\big(\kappa^{4}-3(\kappa')^{2}-4\kappa\kappa''\big)\cos\alpha\\
    &+(6\kappa^{2}\kappa'-\kappa''')\sin\alpha\\
    y''' &=-\kappa^{2}\sin\alpha+\kappa'\cos\alpha\,,\\
    y^{(4)}&=(\kappa''-\kappa^{3})\cos\alpha-3\kappa\kappa'\sin\alpha\,,\\
    y^{(5)}&=\big(\kappa^{4}-3(\kappa')^{2}-4\kappa\kappa''\big)\sin\alpha\\
    &-(6\kappa^{2}\kappa'-\kappa''')\cos\alpha\,.
  \end{split}
\end{equation}
Evaluating \eqref{eqn:xy_derivs} at point $P$ and utilizing \eqref{eqn:param_repr_coeff_def}, we obtain the coefficients given in the theorem.

\end{IEEEproof}

\begin{theorem}\label{thm:gen_curve_func_repr}
  Given a curve in $xy$-plane with known ($x(s)$, $y(s)$, $\alpha(s)$, $\kappa(s)$), its function representation about point $P$ is unique. The coefficients until fifth-order are
    \begin{equation}
      \begin{split}
        \varphi_{0} &= y_{0}\,, \qquad\hspace{50pt}
        \varphi_{1} = \tan\alpha_{0}\,,\\
        \varphi_{2} &= \dfrac{\kappa_{0}}{2\cos^{3}\alpha_{0}}\,, \qquad\qquad
        \varphi_{3} = \dfrac{\kappa_{0}'+3\kappa_{0}^{2}\tan\alpha_{0}}{6\cos^{4}\alpha_{0}}
        \,,\\
        \varphi_{4} &= \dfrac{5\kappa_{0}^3 }{8\cos^{7}\alpha_{0}}
            +\dfrac{\kappa_{0}''-12 \kappa_{0}^3+10 \kappa_{0}\kappa_{0}'\tan\alpha_{0} }{24\cos^{5}\alpha_{0}}  \,,\\
        \varphi_{5} &= \dfrac{7\kappa_{0}^{2} (\kappa_{0}'+\kappa_{0}^{2}\tan\alpha_{0})}{8\cos^{8}\alpha_{0}}
        +\dfrac{\kappa_{0}'''-86 \kappa_{0}^{2} \kappa_{0}'}{120\cos^{6}\alpha_{0}}\\
        &+\dfrac{ \big(3\kappa_{0}\kappa_{0}''+2(\kappa_{0}')^{2}-12\kappa_{0}^{4}\big) \tan\alpha_{0} }{24\cos^{6}\alpha_{0}}
          \,.
      \end{split}
    \end{equation}
\end{theorem}
\begin{IEEEproof}
  Given an arbitrary point $(x, y)$ on the curve (\ref{eqn:EOM_ref_path_ds}), the derivatives of the coordinates $x$ and $y$ with respect to $s$ at that point are given in \eqref{eqn:xy_derivs}. Also, notice that the derivatives of $y$ with respect to $x$ are the same as those given in (\ref{eqn:proof_dyx}, \ref{eqn:proof_dyx_recurv}). Substituting \eqref{eqn:xy_derivs} into (\ref{eqn:proof_dyx}, \ref{eqn:proof_dyx_recurv}), one can obtain the derivatives recursively until the required order. Then the coefficients can be obtained by evaluating these derivatives at point $P$ and utilizing \eqref{eqn:func_repr_coeff_def}.
\end{IEEEproof}

\begin{remark}
One can verify that coefficients given in Theorem~\ref{thm:gen_curve_param_repr} and \ref{thm:gen_curve_func_repr} satisfy the transformations given in Theorem~\ref{thm:curve_func_2_param_repr} and \ref{thm:curve_param_2_func_repr}.
\end{remark}

\section{Application in Lateral Control \label{sec:veh_ctrl_setup}}
This section applies arc-length-based parametric representation to camera-based lateral control problem. We first present a typical architecture using function representation, and discuss the related issues in lane estimation and control. Then a new architecture is proposed to use parametric representation that can facilitate and improve lane estimation as well as control-related information extraction.

\begin{figure}[!t]
  \centering
  \includegraphics[scale=1.2]{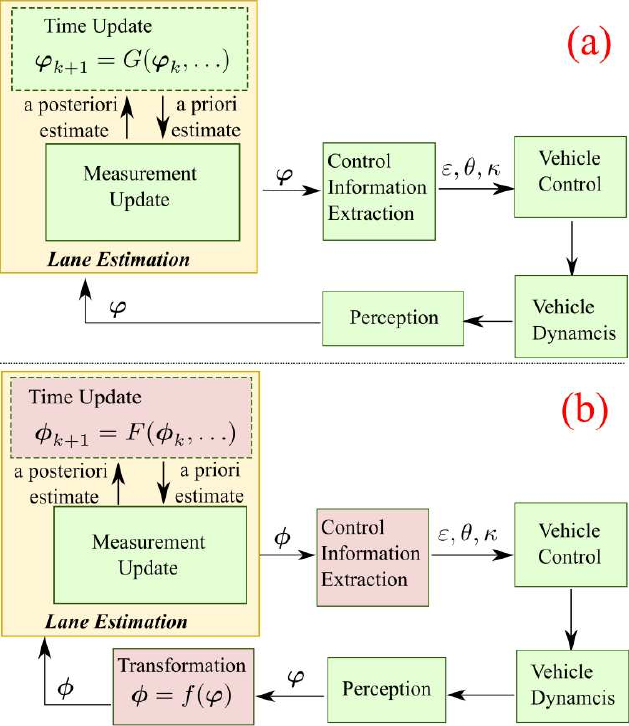}\\
  \caption{Block diagrams of camera-based vehicle control. (a) A typical architecture using function representation. (b) Proposed architecture using arc-length-based parametric representation. }\label{fig:block_diag}
\end{figure}

Fig.~\ref{fig:block_diag}(a) illustrates a typical architecture of camera-based vehicle control. At first, lanes are captured by cameras. Then perception applies lane detection algorithms using computer vision or machine learning techniques, and outputs coefficients $\boldsymbol{\varphi}$ that represent lanes with polynomial function representations. In the scenarios where perception is not perfect (noisy or corrupted) or temporarily unavailable, lane estimation becomes extremely important, which is achieved through predictors, observers or estimators. Those techniques require a model on the evolution of coefficients for prediction. For example, the orange box in Fig.~\ref{fig:block_diag}(a) represents a typical Kalman filter implementation, which is decoupled into two steps: time update and measurement update. Time update step utilizes a model to predict a priori estimate of the coefficients, while the measurement update step generates a posteriori estimate based on the newest measurement and signal characteristics.
However, when function representation is used, it is rather difficult to find such a model because: i) the vehicle body-fixed frame is translating and rotating due to vehicle movement; and ii) function representation does not preserve the form in coordinate transformation.
Therefore, in practice it is common to use simple approximated models for prediction, such as $\boldsymbol{\varphi}_{k+1}=\boldsymbol{\varphi}_{k}$. These predictions only work for a very short period, and underlying inaccuracies in the model lead to unexpected behaviors due to the aforementioned issues.
The other issue appears when control algorithms need to extract state information with respect to lanes at a preview distance, such as lateral deviation, relative heading angle, road curvature, etc. Extracting such information requires numerical iterations for function representation because the preview distance is implicit in the representation.

These issues can be resolved if arc-length-based parametric representation is utilized to characterize lanes.
When the vehicle moves, the nice properties in Theorem~\ref{thm:curve_param_repr_shift} and \ref{thm:curve_param_repr_conformal} indicate an intrinsic linear model on the evolution of coefficients. Also, it is straightforward to extract information with respect to lanes at a preview distance since this distance is actually the arc length parameter explicitly used in the representation.
Fig.~\ref{fig:block_diag}(b) proposes a new architecture for camera-based control problem using arc-length-based parametric representation.
We still assume perception outputs coefficients $\boldsymbol{\varphi}$ of polynomial function representation to ensure compatibility with current platforms. An additional step is introduced to transform function representation to parametric representation by applying Theorem~\ref{thm:curve_func_2_param_repr}. Thus, we can derive an intrinsically linear model on the evolution of coefficients $\boldsymbol{\phi}$, and easily extract information needed by control algorithms. The details are provided in the remainder of this part. In Section~\ref{sec:perception}, we introduce notations, and discuss lane representations and transformations. Section~\ref{sec:estmation_lane} derives the model on the evolution of coefficients $\boldsymbol{\phi}$ for lane estimation. In Section~\ref{sec:ctrl}, we introduce a vehicle dynamic model and a lateral controller as an example to demonstrate how to utilize the derived lane estimation model.

\subsection{Perception and Transformation \label{sec:perception}}

\begin{figure}[!t]
  \centering
  \includegraphics[scale=0.9]{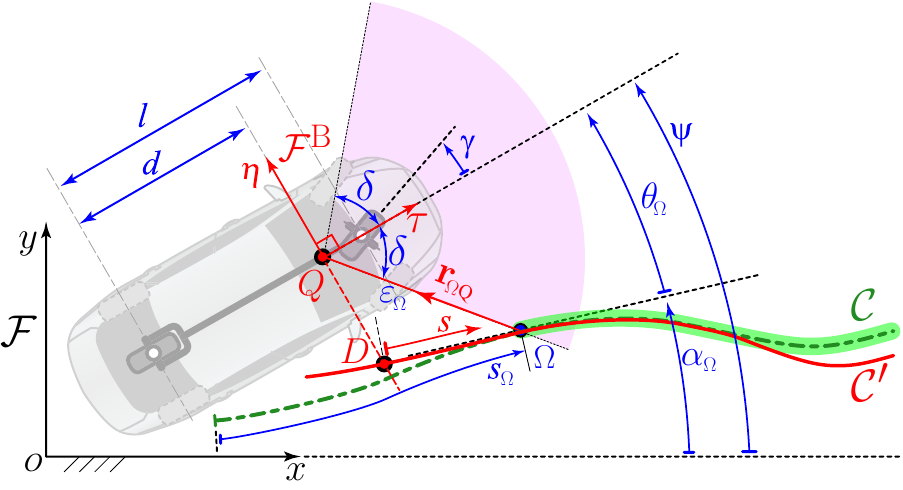}\\
  \caption{Schematics of camera-based vehicle control.\label{fig:camera_control}}
\end{figure}

Fig.~\ref{fig:camera_control} depicts the camera-based perception of the path represented as the green dashed curve. Remark that in practice perception algorithms output representations of lane markers captured by cameras, which can be transformed into representations of lane centers. For simplicity, in this paper we use paths or lanes to indicate lane centers.
We also assume that the camera with FOV angle $2\delta$ is mounted at point $Q$ along the longitudinal symmetry axis. The wheelbase length is $l$, and the distance from point $Q$ to the rear axle center is $d$. The light purple sector region denotes the FOV of the camera, which is also symmetric about the longitudinal axis. The closest on-path point observed in the camera is $\Omega$. Note that the camera can only capture the segment  $\mathcal{C}$ of the path within the FOV that is highlighted as the solid green curve beyond point $\Omega$. In this part, we maintain the following notations:
\begin{enumerate}
  \item ${(x,y,z)}$ denotes the earth-fixed frame $\cFE$.

  \item ${(\tau, \eta, z)}$ denotes the vehicle body-fixed frame $\cFB$ with the origin located at $Q$. Axes $\tau$ and $\eta$ are along the longitudinal and lateral directions, with the corresponding unit vectors denoted as $\unitvec[\tau]$ and $\unitvec[\eta]$, respectively.

  \item The position of $Q$ is ($x_{Q}$, $y_{Q}$) expressed in $\mathcal{F}$, and the vehicle heading angle is $\psi$ with respect to the $x$-axis.

  \item The position of $\Omega$ is ${(x_{\rm \Omega}, y_{\rm \Omega})}$ expressed in $\mathcal{F}$, while the slope angle and curvature at point $\Omega$ are $\alpha_{\rm \Omega}$ and $\kappa_{\rm \Omega}$, respectively.

  \item When it is needed to distinguish different time instants, subscripts $k$ or $k+1$ indicating time steps will be added to the points, axes, frames and time-varying variables.
\end{enumerate}

Perception algorithms process the observed path segment $\mathcal{C}$ and approximate it with a polynomial function
\begin{align}\label{eqn:poly_func_repr_D}
    \mathcal{C}'=\{(\tau, \eta)\,|\,\eta = \boldsymbol{\varphi}(0)\,\boldsymbol{p}_{N}(\tau)\,,\,\tau\in\mathbb{R}\}\,,
\end{align}
that is expressed in frame $\cFB$ and indicated by the red curve in Fig.~\ref{fig:camera_control}.
Note that point $D$ corresponding to $\tau=0$ is invisible in the FOV since it is typically less than $180$ degrees. Strictly speaking, \eqref{eqn:poly_func_repr_D} is not the Taylor approximation of $\mathcal{C}$ about point $D$, but the expansion of Taylor approximation of $\mathcal{C}$ about the closest on-path point (point $\Omega$) in the FOV, that is,
\begin{align}\label{eqn:poly_func_repr_omega}
  \mathcal{C}'=\{(\tau, \eta)\,|\,\eta = \boldsymbol{\varphi}(\tau_{\Omega})\,\boldsymbol{p}_{N}(\tau-\tau_{\Omega})\,,\,\tau\in\mathbb{R}\}\,.
\end{align}
According to Remark~\ref{remark:func_repr_preserve_shift}, the change of coefficients is
\begin{align}
     \boldsymbol{\varphi}^\top(0)&=\boldsymbol{T}(-\tau_{\Omega})\boldsymbol{\varphi}^{\top}(\tau_{\Omega})\,,
\end{align}
where $\boldsymbol{T}(\cdot)$ is the same as that given in \eqref{eqn:T_expansion}.

Next step is to transform function representation \eqref{eqn:poly_func_repr_D} to arc-length-based parametric representation.
Indeed, one should transform \eqref{eqn:poly_func_repr_omega} to parametric representation about point $\Omega$ since \eqref{eqn:poly_func_repr_D} is the expansion of approximation \eqref{eqn:poly_func_repr_omega} about point $\Omega$.
However, it is rather complicated to transform to parametric representation about point $\Omega$; see proof of Theorem~\ref{thm:curve_func_2_param_repr}. Also, transformation about point $\Omega$ leads to the issue of floating origin of the arc length coordinate $s$ while the vehicle is moving. Therefore, we transform \eqref{eqn:poly_func_repr_D} to parametric representation about point $D$, which is the $\eta$-intercept of $\mathcal{C}'$ in frame $\cFB$.
As shown in Fig.~\ref{fig:camera_control}, point $\Omega$ is close to point $D$ when the following holds: i) the lateral deviation to the path is small; or ii) the camera has a relatively wide view. Applying Theorem~\ref{thm:curve_func_2_param_repr}, we obtain the parametric representation of $\mathcal{C}'$ given in \eqref{eqn:poly_func_repr_D} about point $D$ as
\begin{align}\label{eqn:param_repr_body_frame}
  \mathcal{C}'&=\{\bfr\,|\,\bfr(s; 0)= \boldsymbol{\phi}(0)\,\boldsymbol{p}_{N}(s)\,,\,s\in\mathbb{R}\}\,,
\end{align}
where $\bfr = [\tau, \; \eta]^\top$ is the coordinates in frame $\cFB$, the location of arc length coordinate $s=0$ corresponds to point $D$, and the change of coefficients is
\begin{align}\label{eqn:param_repr_body_frame_coeff}
  \boldsymbol{\phi}(0)&= \boldsymbol{f}(\boldsymbol{\varphi}(0))\,.
\end{align}

\subsection{Lane Estimation \label{sec:estmation_lane}}

\begin{figure}
  \centering
  \includegraphics[scale=0.6]{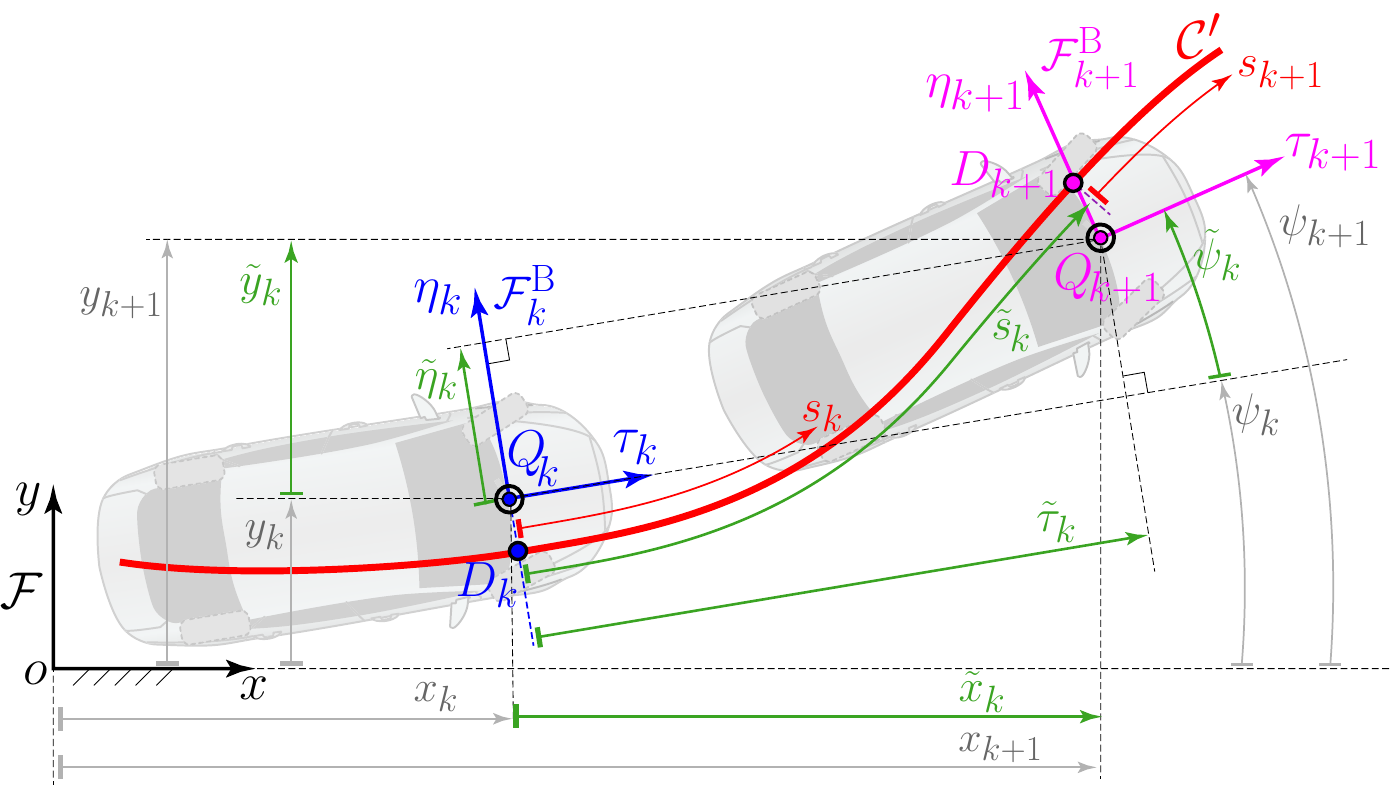}\\
  \caption{Vehicle movement from step $k$ to $k+1$.\label{fig:lanepoly_time_evolve}}
\end{figure}

Fig.~\ref{fig:lanepoly_time_evolve} depicts vehicle movement from time step $k$ to $k+1$. As mentioned earlier, Fig.~\ref{fig:lanepoly_time_evolve} uses subscripts $k$ or $k+1$ on points, frames and time-varying variables to distinguish different instants; cf.~Fig.~\ref{fig:camera_control}. Specifically, point $Q_{k}$, frame $\cFB[k]$ with $\tau_{k}\eta_{k}z$ axis and point $D_{k}$ denote the location of the camera, the body-fixed frame and the $\eta$-intercept of the lane at step $k$, respectively. The red curve $\mathcal{C}'$ represents the curve \eqref{eqn:param_repr_body_frame} at step $k$.
$(x_{k}, y_{k})$ are the coordinates of $Q_{k}$ in the earth-fixed frame $\cFE$, while $\psi_{k}$ is the vehicle heading. $(\tilde{x}_{k}, \tilde{y}_{k})$ and $(\tilde{\tau}_{k}, \tilde{\eta}_{k})$ are the coordinates of the displacement vector $Q_{k}Q_{k+1}$ expressed in the earth-fixed frame $\cFE$ and body-fixed frame $\cFB[k]$, respectively. $\tilde{\psi}_{k}$ is the change of vehicle heading from step $k$ to step $k+1$. $s_{k}$ is the arc length coordinate at step $k$ such that $s_{k}=0$ corresponds to point $D_{k}$, while $\tilde{s}_{k}$ represents the arc length distance from $D_{k}$ to $D_{k+1}$ along the curve $\mathcal{C}'$. In summary,
\begin{equation}\label{eqn:term_diff_k_to_k1}
\begin{split}
  \tilde{x}_{k} & = x_{k+1}-x_{k}\,,\qquad
  \tilde{y}_{k} = y_{k+1}-y_{k}\,,\\
  \tilde{\psi}_{k} & = \psi_{k+1}-\psi_{k}\,,\qquad
  \tilde{s}_{k} = s_{k}-s_{k+1}\,.
\end{split}
\end{equation}
In the following we assume vehicle state changes $\tilde{x}_{k}$, $\tilde{y}_{k}$, $\tilde{\psi}_{k}$, $\tilde{\tau}_{k}$, $\tilde{\eta}_{k}$ and $\tilde{s}_{k}$ are known, and will investigate the details on how to obtain them in Section~\ref{sec:ctrl}.

To derive the evolution of coefficients in \eqref{eqn:param_repr_body_frame}, we highlight the time-dependency and rewrite it for step $k$ as
\begin{align}\label{eqn:param_repr_body_frame_k_in_k}
  \mathcal{C}'&=\{\bfr_{k}\,|\,\bfr_{k}(s_{k}; 0) = \boldsymbol{\phi}_{k}(0)\,\boldsymbol{p}_{N}(s_{k})\,,\,s_{k}\in \mathbb{R\}}\,,
\end{align}
which is the parametric representation using arc length coordinates $s_{k}$ about point $D_{k}$ in frame $\cFB[k]$. The objective is to derive the new representation
\begin{equation}\label{eqn:param_repr_body_frame_k+1}
\begin{split}
    \mathcal{C}'&=\{\bfr_{k+1}\,|\,
  \bfr_{k+1}(s_{k+1}; 0)= \boldsymbol{\phi}_{k+1}(0)\,\boldsymbol{p}_{N}(s_{k+1})\,,\, s_{k+1}\in\mathbb{R}\}\,,
\end{split}
\end{equation}
given \eqref{eqn:param_repr_body_frame_k_in_k} based on vehicle state changes from step $k$ to step $k+1$ such that the relationship between the coefficients $\boldsymbol{\phi}_{k}(0)$ and $\boldsymbol{\phi}_{k+1}(0)$ can be obtained. Note that \eqref{eqn:param_repr_body_frame_k+1} is based on arc length coordinates $s_{k+1}$ about point $D_{k+1}$ in frame $\cFB[k+1]$. Hence, three steps are performed sequentially in the following: i) coordinate transformation from frame $\cFB[k]$ to frame $\cFB[k+1]$; ii) shifting expansion point from $D_{k}$ to $D_{k+1}$; iii) changing coordinate from $s_{k}$ to $s_{k+1}$.

According to Theorem~\ref{thm:gen_coord_transf}, the change of coordinates from body-fixed frame $\cFB[k+1]$ to $\cFB[k]$ is
\begin{align}\label{eqn:coord_transf_k_k_1}
\boldsymbol{r}_{k} &= \bfR_{k}\, \boldsymbol{r}_{k+1}+\boldsymbol{d}_{k}\,,
\end{align}
where
\begin{align}
    \bfR_{k} &=
    \begin{bmatrix}
      \cos\tilde{\psi}_{k}  & -\sin\tilde{\psi}_{k}  \\ \sin\tilde{\psi}_{k}  &\cos\tilde{\psi}_{k}
    \end{bmatrix}\,,&
    \bfd_{k} &=
    \begin{bmatrix}
      \tilde{\tau}_{k}   \\ \tilde{\eta}_{k}
    \end{bmatrix}\,.
\end{align}
Applying Theorem~\ref{thm:curve_param_repr_conformal} to curve \eqref{eqn:param_repr_body_frame_k_in_k} with coordinate transformation \eqref{eqn:coord_transf_k_k_1}, we obtain the parametric representation using arc length coordinates $s_{k}$ about point $D_{k}$ in frame $\cFB[k+1]$ as
\begin{align}\label{eqn:param_repr_body_frame_k_in_k+1}
  \mathcal{C}'&=\{\bfr_{k+1}\,|\,
  \bfr_{k+1}(s_{k}; 0)= \widehat{\boldsymbol{\phi}}_{k}(0)\,\boldsymbol{p}_{N}(s_{k})\,,\,s_{k}\in\mathbb{R}\}\,,
\end{align}
where the change of coefficients is
\begin{align}\label{eqn:coeff_change_step_1}
  \widehat{\boldsymbol{\phi}}_{k}(0) & = \bfR_{k}^{\top}\big(\boldsymbol{\phi}_{k}(0)-\bfD_{k} \big)\,,
\end{align}
and
\begin{align}
\bfD_{k} & = \begin{bmatrix}
  \bfd_{k} & 0 & \cdots & 0
\end{bmatrix}\in \mathbb{R}^{2\times (N+1)}\,.
\end{align}

Notice that the expansion point has shifted from $D_{k}$ to $D_{k+1}$ with distance $\tilde{s}_{k}$.
Applying Theorem~\ref{thm:curve_param_repr_shift} to \eqref{eqn:param_repr_body_frame_k_in_k+1}, we obtain the parametric representation of $\mathcal{C}'$ using arc length coordinates $s_{k}$ about point $D_{k+1}$ in frame $\cFB[k+1]$ as
\begin{align}\label{eqn:param_repr_body_frame_k_in_k+1_sk}
 \mathcal{C}'&=\{\bfr_{k+1}\,|\,
  \bfr_{k+1}(s_{k}; \tilde{s}_{k}) = \widehat{\boldsymbol{\phi}}_{k}(\tilde{s}_{k})\,\boldsymbol{p}_{N}(s_{k}-\tilde{s}_{k})\,,\,s_{k}\in\mathbb{R}\}\,,
\end{align}
where the change of coefficients is
\begin{align}\label{eqn:coeff_change_step_2}
  \widehat{\boldsymbol{\phi}}_{k}^\top(\tilde{s}_{k})&=\boldsymbol{T}(\tilde{s}_{k})\,\widehat{\boldsymbol{\phi}}_{k}^\top(0)\,,
\end{align}
where $\boldsymbol{T}(\cdot)$ is the same as that given in \eqref{eqn:T_expansion}.

Next utilizing \eqref{eqn:term_diff_k_to_k1}, we can change coordinates from $s_{k}$ to $s_{k+1}$ to obtain the parametric representation \eqref{eqn:param_repr_body_frame_k+1} where the change of coefficients is
\begin{align}\label{eqn:coeff_change_step_3}
  \boldsymbol{\phi}_{k+1}(0)&=\widehat{\boldsymbol{\phi}}_{k}(\tilde{s}_{k})\,.
\end{align}
In summary, the evolution of coefficients in the parametric representation from step $k$ to step $k+1$ is
\begin{align}\label{eqn:coeff_dyn_gen_equiv}
  \boldsymbol{\phi}_{k+1}(0)&=\bfR_{k}^{\top}\big(\boldsymbol{\phi}_{k}(0)-\bfD_{k} \big)\, \boldsymbol{T}^\top(\tilde{s}_{k})\,,
\end{align}
cf. (\ref{eqn:coeff_change_step_1}, \ref{eqn:coeff_change_step_2}, \ref{eqn:coeff_change_step_3}), implying a linear relationship by nature.

\begin{remark}
Utilizing vectorization operator and Theorem~\ref{thm:vec_kron_relation}, one can rewrite \eqref{eqn:coeff_dyn_gen_equiv} into the standard form of linear system
\begin{align}
  \boldsymbol{\Phi}_{k+1}&=\mathbf{A}_{k}\boldsymbol{\Phi}_{k}+\mathbf{B}_{k}\,,
\end{align}
where
\begin{equation}
  \begin{split}
    \boldsymbol{\Phi}_{k}&=\vectr(\boldsymbol{\phi}_{k}(0))\,, \qquad\qquad
   \mathbf{A}_{k} =\boldsymbol{T}(\tilde{s}_{k})\otimes \bfR_{k}^{\top}\,,\\
   \mathbf{B}_{k} &= -\big(\boldsymbol{T}(\tilde{s}_{k})\otimes \bfR_{k}^{\top}\big)\vectr(\mathbf{D}_{k})\,.
  \end{split}
\end{equation}
\end{remark}

\subsection{Vehicle Dynamics and Control \label{sec:ctrl}}

The proposed architecture and model on the evolution of coefficients using arc-length-based parametric representation can be applied to any vehicle models with reasonable control algorithms that require lane estimation and information extraction. To demonstrate the work flow, in this part we provide an example on the model derived in \cite{Wubing_ND_2022} using the path-following controller proposed in \cite{Wubing_LC_TIV_2022}.

\subsubsection{Dynamics and Transformation on States\label{sec:dyn_estimation_states}}

In general, vehicle dynamics describe the evolution of absolute position (e.g., $x_{\rm Q},\, y_{\rm Q}$) and orientation (heading angle $\psi$) in the earth-fixed frame $\cFE$. They have different levels of complexity based on fidelity.
Here, we consider the model derived in \cite{Wubing_ND_2022} on the camera location point $Q$, that is
\begin{equation}\label{eqn:EOM_point_Q}
\begin{split}
\dot{x}_{\rm Q} &= V\,\cos\psi-d\,\dfrac{V}{l}\tan\gamma \sin\psi\, , \\
\dot{y}_{\rm Q} &= V\,\sin\psi+d\,\dfrac{V}{l}\tan\gamma \cos\psi\, , \\
\dot{\psi} &= \dfrac{V}{l}\,\tan\gamma\, .
\end{split}
\end{equation}
where $x_{\rm Q}$, $y_{\rm Q}$, ${\psi}$, $l$ and $d$ utilize the same notations introduced in Section~\ref{sec:perception}, $V$ is the constant longitudinal speed, and $\gamma$ is the steering angle.

To facilitate control design, the vehicle dynamics \eqref{eqn:EOM_point_Q} on absolute position ($x_{\rm Q},\, y_{\rm Q}$) and orientation (heading angle $\psi$) expressed in the earth-fixed frame $\cFE$ can be transformed to dynamics on relative position and orientation with respect to the path. Such transformation allows us to obtain the evolution of arc length position, lateral deviation and relative heading with respect to lanes perceived in the camera. In this paper, we choose the following relative position and orientation as new states: i) the arc length $s_{\Omega}$ that the vehicle has travelled along the lane; ii) the observed distance $\varepsilon_{\Omega}$ that characterizes the length of vector $\pvec{\Omega Q}$ (cf.~Fig.~\ref{fig:camera_control}), whose sign is positive when $Q$ is on the left side of the path;
and iii) the relative heading angle
\begin{align}\label{eqn:rel_heading_def}
  \theta_{\Omega} &= \psi-\alpha_{\Omega}\,,
\end{align}
with respect to point $\Omega$ on the path.
The details on the state transformation are provided in Appendix~\ref{append:coord_transf}, and the resulting transformed relative dynamics are
\begin{equation}\label{eqn:transf_xy_se_diff_Q}
  \begin{split}
    \dot{s}_{\Omega} &= \dfrac{ V }{\sin(\delta-\theta_{\Omega})}\big(\sin\delta+\dfrac{d}{l}\cos\delta \tan\gamma\big)\\
    &+\dfrac{|\varepsilon_{\Omega}|\cos\big(\delta-\sign(\varepsilon_{\Omega})\,\delta\big)}{\sin(\delta-\theta_{\Omega})}\dfrac{V}{ l}\tan\gamma\,,\\
    \dot{\varepsilon}_{\Omega} &=\dfrac{ V }{\sin(\delta-\theta_{\Omega})}\big(\sin\theta_{\Omega}+\dfrac{d}{l}\cos\theta_{\Omega} \tan\gamma\big)\\
    &+\dfrac{|\varepsilon_{\Omega}|\cos(\delta-\sign(\varepsilon_{\Omega})\,\theta_{\Omega})}{\sin(\delta-\theta_{\Omega})}\dfrac{V}{l}\tan\gamma\,,\\
    \dot{\theta}_{\Omega} &=-\dfrac{ V\, \kappa_{\Omega}}{\sin(\delta-\theta_{\Omega})}\big(\sin\delta+\dfrac{d}{l}\cos\delta \tan\gamma\big)\\
    &+\Big(1-\dfrac{\kappa_{\Omega} |\varepsilon_{\Omega}|\cos\big(\delta-\sign(\varepsilon_{\Omega})\,\delta\big)}{\sin(\delta-\theta_{\Omega})}\Big)\dfrac{V}{l}\tan\gamma\,.
  \end{split}
\end{equation}

\subsubsection{Control}

The objective of a path-following controller is to generate desired steering angle $\gamma_{\rm des}$ such that point Q can follow the given path. In literature, there are many available controllers demonstrated to be effective in different scenarios. In this part we use the nonlinear controller proposed in \cite{Wubing_LC_TIV_2022} as an example since cameras are typically mounted at the front of the vehicle.
We hightlight the key ideas of this controller, and refer readers to \cite{Wubing_LC_TIV_2022} for more details on the design.

With the assumption that the steering angle $\gamma$ can track any desired value $\gamma_{\rm des}$, that is, $\gamma = \gamma_{\rm des}$,
the path-following controller is
\begin{align}\label{eqn:lateral_controller}
  \gamma_{\rm des} & = \gamma_{\rm ff} + \gamma_{\rm fb}\,,
\end{align}
which consists of a feedforward control law
\begin{align}
  \gamma_{\rm ff} &= \arctan \dfrac{l\,\kappa_{\rm D}}{\sqrt{1-(d\,\kappa_{\rm D})^{2}}}\,, \label{eqn:steer_controller_ff_sum}
\end{align}
and a feedback control law
\begin{align}\label{eqn:steer_controller_fb_sum}
  \gamma_{\rm fb} &= \gamma_{\rm sat}\cdot g \Big( \tfrac{k_{1}}{\gamma_{\rm sat}}\big(\theta_{\rm D}-\theta_{0}+\arctan (k_{2}\,\varepsilon_{\rm D})\big)\Big)\,.
\end{align}
Here, $\kappa_{\rm D}$ is the road curvature at point $D$, $\theta_{\rm D}$ is the relative heading angle with respect to point $D$ (cf.~\eqref{eqn:rel_heading_def}), and ${\varepsilon_{\rm D}=-\eta_{\rm D}}$ is the $\eta$-deviation of point $D$ where $\eta_{\rm D}$ indicates the $\eta$-coordinate of point $D$ in frame $\cFB$. In other words, $\varepsilon_{\rm D}$ characterizes the length of vector $\pvec{D Q}$ (cf.~Fig.~\ref{fig:camera_control}), and is positive when $\pvec{D Q}$ points towards the positive $\eta$-axis.
Also,
\begin{align}\label{eqn:theta0_des}
  \theta_{0} & = -\arcsin(d\,\kappa_{\rm D})\,
\end{align}
is the desired yaw angle error, ($k_{1}$, $k_{2}$) are tunable control gains, $\gamma_{\rm sat}$ is the maximum allowable steering angle, and $g (x)$ denotes the wrapper function
\begin{equation}
    g(x) =\dfrac{2}{\pi}\arctan \Big(\dfrac{\pi}{2} x\Big)\,. \label{eqn:satfunction}
\end{equation}
The feedforward control essentially provides the estimated steering angle to handle a given road curvature $\kappa_{\rm D}$ whereas the feedback control makes corrections based on lateral deviation $\varepsilon_{\rm D}$ and yaw angle error $\theta_{\rm D}-\theta_{0}$.

The controller (\ref{eqn:lateral_controller}-\ref{eqn:satfunction}) relies on the information about point $D$ or with respect to point $D$. We remark that point $D$ is not the closest on-path point to $Q$ as that used in \cite{Wubing_LC_TIV_2022}, or the closest observable on-path point (point $\Omega$) to $Q$ in the FOV. One can verify that when the road curvature is constant (i.e., $\kappa_{\rm D}=\kappa^{\ast}$), the closed loop system (\ref{eqn:transf_xy_se_diff_Q}-\ref{eqn:satfunction}) possesses the desired equilibrium
\begin{align}\label{eqn:equilb}
  s_{\Omega}^{\ast}=\dfrac{V\,t}{\sqrt{1-(d\,\kappa^{\ast})^{2}}}\,,\;
  \varepsilon_{\Omega}^{\ast}=0\,,\;
  \theta_{\Omega}^{\ast}=-\arcsin(d\,\kappa^{\ast})\,,
\end{align}
which can be stabilized with properly chosen gain $k_{1}$ and $k_{2}$. The equilibrium \eqref{eqn:equilb} represents the scenario where the vehicle follows the given path perfectly with no lateral deviations.
To extract the information ($\kappa_{\rm D}$, $\theta_{\rm D}$, $\varepsilon_{\rm D}$) from representation \eqref{eqn:param_repr_body_frame}, we provide the following lemma.


\begin{lemma}\label{thm:terms_from_param_repr}
  Given the parametric representation \eqref{eqn:param_repr_body_frame} of the lane in frame $\cFB$, the $\eta$-deviation $\varepsilon_{\rm D}$, the relative heading angle $\theta_{\rm D}$, and the curvature $\kappa_{\rm D}$ at point $D$ are
  \begin{equation}\label{eqn:extract_ctrl_info_from_repr}
    \begin{split}
      \varepsilon_{D} &=-\hat{\phi}_{0}\,,\,\qquad\qquad\qquad
      \theta_{D} = -\arctan \dfrac{\hat{\phi}_{1}}{\bar{\phi}_{1}}\,,\,\\
      \kappa_{D} &= \dfrac{2\hat{\phi}_{2}\bar{\phi}_{1}-2\hat{\phi}_{1}\bar{\phi}_{2}}{\hat{\phi}_{1}^{2}+\bar{\phi}_{1}^{2}}\,.
    \end{split}
  \end{equation}
\end{lemma}
\begin{IEEEproof}
  Given \eqref{eqn:param_repr_body_frame}, we obtain the $\eta$-deviation $\varepsilon$, the relative heading angle $\theta$, and the road curvature $\kappa$ at an arbitrary point $P$ with a preview distance $s_{\rm P}$ as
  \begin{equation}\label{eqn:lemma_get_terms_from_repr}
    \begin{split}
      \varepsilon(s_{\rm P}) &= -\eta(s_{\rm P}) \,,\qquad
      \theta(s_{\rm P}) = -\arctan \dfrac{\eta'(s_{\rm P})}{\tau'(s_{\rm P})}\,,\\
      \kappa(s_{\rm P}) &= \dfrac{\eta''(s_{\rm P})\,\tau'(s_{\rm P})-\eta'(s_{\rm P})\,\tau''(s_{\rm P})}{(\eta'(s_{\rm P}))^{2}+(\tau'(s_{\rm P}))^{2}}\,,
    \end{split}
  \end{equation}
  where the derivatives $\eta'$, $\tau'$, $\eta''$ and $\tau''$ can be obtained by differentiating \eqref{eqn:param_repr_body_frame} with respect to $s$.
  Point D marks the origin of arc length coordinate $s$. Thus, evaluating \eqref{eqn:lemma_get_terms_from_repr} at $s_{\rm P}=0$ yields \eqref{eqn:extract_ctrl_info_from_repr} at point $D$.
\end{IEEEproof}

\subsubsection{Vehicle State Changes}
In the derivation of \eqref{eqn:coeff_dyn_gen_equiv}, vehicle state changes $\tilde{x}_{k}$, $\tilde{y}_{k}$, $\tilde{\psi}_{k}$, $\tilde{\tau}_{k}$, $\tilde{\eta}_{k}$ and $\tilde{s}_{k}$ are assumed to be known at each step in Section~\ref{sec:estmation_lane}. In practice, they can be obtained based on a nominal vehicle dynamic model and sensor data. In this part we take model \eqref{eqn:EOM_point_Q} as an example, and assume the longitudinal speed $V$ and yaw rate $\omega=\dot{\psi}$ can be measured by onboard sensors. We rewrite model \eqref{eqn:EOM_point_Q} into
\begin{equation}\label{eqn:model_w_sensor_data}
  \begin{split}
    \dot{x}&= V\cos\psi -d\, \omega \sin\psi\, ,\\
    \dot{y}&= V\sin\psi+d\, \omega\cos\psi\,, \\
    \dot{\psi}&=\omega\,.
  \end{split}
\end{equation}
based on measurable data $V$ and $\omega$.
Applying Euler method to integrate \eqref{eqn:model_w_sensor_data} at step $k$, we obtain
\begin{equation}\label{eqn:changes_x_y_psi}
  \begin{split}
    \tilde{x}_{k}&= (V_{k}\cos\psi_{k} -d\, \omega_{k} \sin\psi_{k})T\,,\\
    \tilde{y}_{k}&= (V_{k}\sin\psi_{k}+d\, \omega_{k}\cos\psi_{k})T\,,\\
    \tilde{\psi}_{k}&=\omega_{k} T\,,
  \end{split}
\end{equation}
where $T$ is the step size. Notice that $(\tilde{x}_{k}, \tilde{y}_{k})$ and $(\tilde{\tau}_{k}, \tilde{\eta}_{k})$ are the coordinates of displacement vector $Q_{k}Q_{k+1}$ expressed in earth-fixed frame $\cFE$ and body-fixed frame $\cFB[k]$, respectively. Thus, one can apply Theorem~\ref{thm:gen_coord_transf} on coordinate transformation and obtain
\begin{align}\label{eqn:changes_tau_eta_psi}
\tilde{\tau}_{k}&= V_{k}T\,,&
\tilde{\eta}_{k}&= d\, \omega_{k} T\,.
\end{align}

Note that $\tilde{s}_{k}$ is the arc length distance that point $D$ has travelled along the curve $\mathcal{C}'$ from step $k$ to step $k+1$. Realizing that point $D$ coincides with point $\Omega$ if $\delta=\frac{\pi}{2}$, we can easily derive the evolution of $s_{\rm D}$ from the relative dynamics \eqref{eqn:transf_xy_se_diff_Q} that is transformed from \eqref{eqn:EOM_point_Q}.
By setting $\delta=\frac{\pi}{2}$ in \eqref{eqn:transf_xy_se_diff_Q} and replacing point $\Omega$ with point $D$, the evolution of $s_{\rm D}$ becomes
\begin{equation}
  \begin{split}\label{eqn:transf_xy_se_diff_D}
    \dot{s}_{\rm D} &= \dfrac{ V+\omega\,\varepsilon_{\rm D} }{\cos\theta_{\rm D}}\,,
  \end{split}
\end{equation}
where yaw rate $\omega=\tfrac{V}{l}\,\tan\gamma$ is utilized; cf.~\eqref{eqn:model_w_sensor_data}.
Applying Lemma~\ref{thm:terms_from_param_repr} to the parametric representation \eqref{eqn:param_repr_body_frame},
the $\eta$-deviation $\varepsilon_{D}$ and relative heading angle $\theta_{D}$ can be obtained at step $k$. Thus, integrating \eqref{eqn:transf_xy_se_diff_D} with Euler method yields
\begin{equation}\label{eqn:changes_s}
  \tilde{s}_{k}=\dfrac{V_{k}+\omega_{k}\,\varepsilon_{\textrm{D}_{k}}}{\cos(\theta_{\textrm{D}_{k}})}\,T\,.
\end{equation}


\section{Results \label{sec:res}}
In this section, simulations of the whole process explained in Section~\ref{sec:veh_ctrl_setup} are performed to demonstrate the usage of arc-length-based parametric representation in lane estimation and vehicle control. In Section~\ref{sec:sim_setup}, we start with the details on how we set up experiment scenarios and simulate lane perception in camera-based control. In Section~\ref{sec:sim_res}, simulations with or without using model \eqref{eqn:coeff_dyn_gen_equiv} for prediction are conducted for the path-following control problem and results indicate the efficacy and large potential of \eqref{eqn:coeff_dyn_gen_equiv} in facilitating lane estimation in vehicle control.

\subsection{Simulation Setup\label{sec:sim_setup}}

The model \eqref{eqn:coeff_dyn_gen_equiv} integrated with the controller (\ref{eqn:lateral_controller}-\ref{eqn:satfunction}) is applicable to lane estimation in path-following problems where the path can have any reasonable shape. Remark~\ref{remark:curv_repr_gen} indicates that a path can be fully described by $\kappa(s)$. Also, if $\kappa(s)$ has an analytical form, we can derive the coefficients for polynomial function representation \eqref{eqn:poly_func_repr_D} given vehicle state with respect to the path. These coefficients can be used as: i) the output of perception algorithms at perception updating instants; and ii) the true values to compare against the predicted values when perception outcome is not available. Therefore, we consider the closed path used in \cite{Wubing_ND_2022} and derive the coefficients for polynomial function representation in this part.
The curvature of the path is given as
\begin{align}\label{eqn:curv_s_func}
 \kappa(s) &= \dfrac{\kappa_{\max}}{2}\left(1-\cos\bigg(\dfrac{2\pi}{s_{\rm T}}s\bigg)\right)\ ,
\end{align}
where $\kappa_{\max}$ is the maximum curvature along the path, and $s_{\rm T}$ is the arc length period.
This path has $N$ corners and perimeter $Ns_{\rm T}$ if
\begin{equation}\label{eqn:curv_close_cond}
 \kappa_{\max}\,s_{\rm T}=\dfrac{4\pi}{N}\ , \quad N=2, 3, \ldots \ .
\end{equation}

As discussed in Section~\ref{sec:perception}, perception algorithms provide representation \eqref{eqn:poly_func_repr_D} of the lane in real-time, which can be viewed as the expansion of Taylor approximation \eqref{eqn:poly_func_repr_omega} about point $\Omega$ in the FOV.
To simulate perception, we need to derive the coefficients of function representation \eqref{eqn:poly_func_repr_omega} in frame $\cFB$ given the vehicle states and the path information \eqref{eqn:curv_s_func}.

\begin{lemma}\label{lemma:gen_curve_func_repr}
    Given the current vehicle relative states ($s_{\Omega}$, $\varepsilon_{\Omega}$ and $\theta_{\Omega}$) with respect to the path, and the half angle $\delta$ of camera FOV, the lane segment $\mathcal{C}$  can be approximated as function representation \eqref{eqn:poly_func_repr_omega} in frame $\cFB$ by applying Taylor series to point $\Omega$, where $\tau_{\Omega}:=\tau(s_{\Omega})=|\varepsilon_{\Omega}| \cos\delta$ and
    \begin{equation}\label{eqn:gen_curv_func_repr_coeff}
      \begin{split}
        \varphi_{0}(\tau_{\Omega}) &= -\varepsilon_{\Omega}\sin\delta\,, \quad
        \varphi_{1}(\tau_{\Omega}) = -\tan\theta_{\Omega}\,, \\
        \varphi_{2}(\tau_{\Omega}) &= \dfrac{\kappa_{\Omega}}{2\cos^{3}\theta_{\Omega}}\,,\quad
        \varphi_{3}(\tau_{\Omega}) = \dfrac{\kappa'_{\Omega}-3\kappa_{\Omega}^{2}\tan\theta_{\Omega}}{6\cos^{4}\theta_{\Omega}}\,,\\
        \varphi_{4}(\tau_{\Omega}) &= \dfrac{5\kappa_{\Omega}^3}{8\cos^{7}\theta_{\Omega}}
        +\dfrac{\kappa''_{\Omega}-12 \kappa_{\Omega}^{3}-10\kappa_{\Omega}\kappa'_{\Omega}\tan\theta_{\Omega} }{24\cos^{5}\theta_{\Omega}}  \,,\\
        \varphi_{5}(\tau_{\Omega}) &=
        \dfrac{ 7\kappa_{\Omega}^{2} (\kappa'_{\Omega}-\kappa_{\Omega}^{2}\tan \theta_{\Omega}) }{8\cos^{8}\theta_{\Omega}}
        +\dfrac{\kappa'''_{\Omega}-86\kappa_{\Omega}^{2} \kappa'_{\Omega}}{120\cos^{6}\theta_{\Omega}} \\
        &+\dfrac{\big(12\kappa_{\Omega}^{4}-3\kappa_{\Omega}\kappa''_{\Omega}-2(\kappa'_{\Omega})^{2}\big)\tan\theta_{\Omega}}{24\cos^{6}\theta_{\Omega}}\,.
      \end{split}
    \end{equation}
\end{lemma}
\begin{IEEEproof}
  See Appendix~\ref{append:path_func_repr_body_frame}.
\end{IEEEproof}

\subsection{Simulation Results \label{sec:sim_res}}

In this part, we simulate the whole process of perception, estimation and control explained in Section~\ref{sec:veh_ctrl_setup} when the vehicle tries to follow the given path (\ref{eqn:EOM_ref_path_ds}, \ref{eqn:curv_s_func}) with camera-based perception. The controller (\ref{eqn:lateral_controller}-\ref{eqn:satfunction}) updates commands every $T$ seconds while perception provides the function representation \eqref{eqn:poly_func_repr_D} every $T_{\rm p}$ seconds. To demonstrate the efficacy of the proposed estimation algorithm, we also assume: i) $T_{\rm p}$ is larger than $T$; ii) when the representation \eqref{eqn:poly_func_repr_D} is not updated by perception, the controller may use the model \eqref{eqn:coeff_dyn_gen_equiv} to update the parametric representation with the estimated state changes (\ref{eqn:changes_x_y_psi}, \ref{eqn:changes_tau_eta_psi}, \ref{eqn:changes_s}) at each step; iii) the measurement \eqref{eqn:poly_func_repr_D} is free of noise such that prediction using model \eqref{eqn:coeff_dyn_gen_equiv} can be easily compared against true values without noise effects; and iv) the true values about coefficients of representation are obtained by calculating $\boldsymbol{\varphi}(\tau_{\Omega})$ using Lemma~\ref{lemma:gen_curve_func_repr} at every control period with vehicle states ($s_{\rm \Omega}$, $\varepsilon_{\Omega}$, $\theta_{\Omega}$). The initial conditions are set to $s_{\rm \Omega}=0$ [m], $\varepsilon_{\Omega}=0.1$ [m], $\theta_{\Omega}=0$ [deg],  and the other parameters used in the simulations are provided in TABLE~\ref{tab:params}.

\begin{table}[!t]
\begin{center}
\renewcommand{\arraystretch}{1.3}
\rowcolors{1}{LightCyan}{LightMagenta}
\begin{tabular}{l|c|l}
\hline\hline
 \rowcolor{Gray}  Parameter & Value & Description\\
 \hline 
 $l$ [m]& $2.57$ & wheelbase length\\
 $d$ [m]& $2$ & distance from Q to rear axle center\\
 $\delta$ [deg]& $60$ & half angle of camera FOV\\
 $k_{1}$ [m/s]& $-l/d$ & control gain\\
 $k_{2}$ [m$^{-1}$]& $0.02$ & control gain\\
 $\gamma_{\max}$ [deg]& $30$ & physical steering angle limit\\
 $s_{\rm T}$ [m] & $250$ & arc length period of the path\\
 $N$ [1] & $4$ & number of corners for closed path\\
 $\kappa_{\max}$ [m$^{-1}$] & $0.004\pi$ & maximum curvature of the path\\
 $V$ [m/s] & 20 & longitudinal speed \\
\hline\hline
\end{tabular}
\end{center}
\caption{Parameters used in the simulations. \label{tab:params}}
\end{table}

\begin{figure}
  \centering
  \includegraphics[scale=1.05]{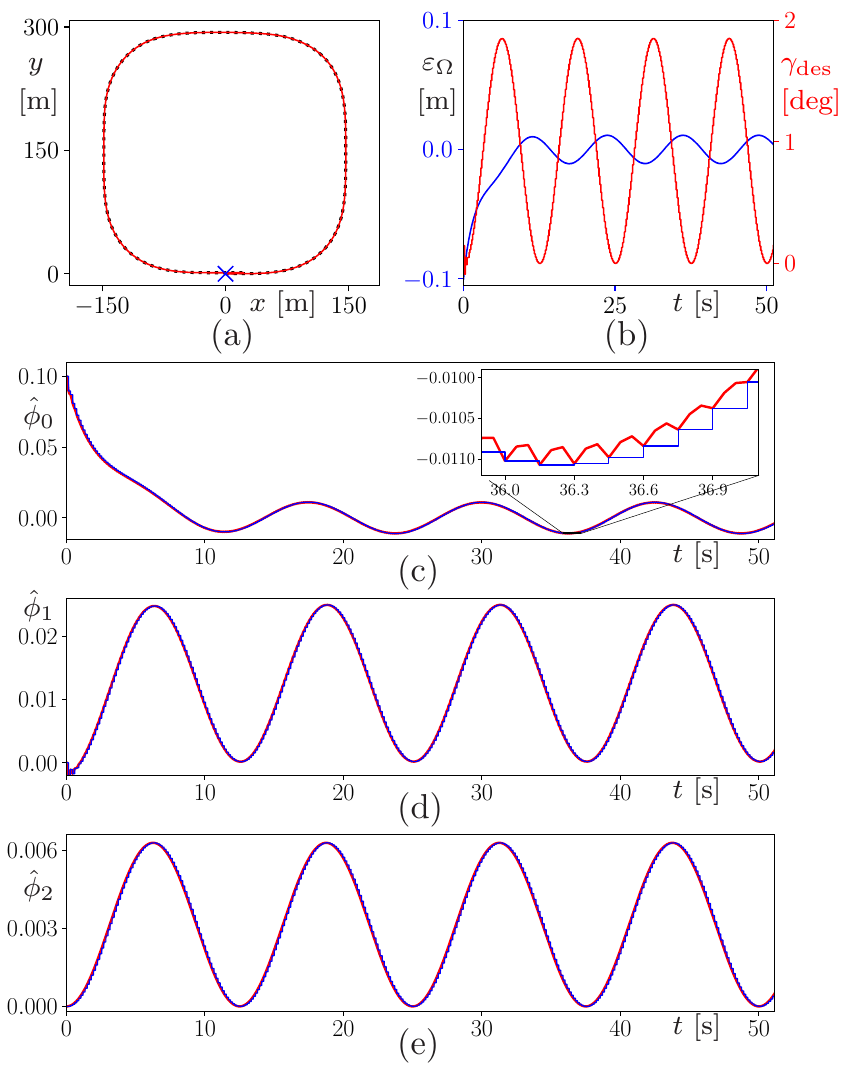}\\
  \caption{Simulation results for camera-based lateral control without prediction on lane representation when $T=0.05$ [s] and $T_{\rm p}=0.15$ [s]. (a) Vehicle position and the closed path (\ref{eqn:EOM_ref_path_ds}, \ref{eqn:curv_s_func}) in ($x$, $y$)-plane. (b) lateral deviation $\varepsilon_{\Omega}$ and steering command $\gamma_{\rm des}$. (c, d, e) time profiles of coefficients in the parametric representation \eqref{eqn:param_repr_body_frame} while the vehicle moves along the path. \label{fig:sim_50ms}}
\end{figure}

Fig.~\ref{fig:sim_50ms} shows the simulation results without using \eqref{eqn:coeff_dyn_gen_equiv} to predict coefficients when $T=0.05$ [s] and $T_{\rm p}=0.15$ [s]. That is, the controller outputs a new command at one step based on the updated lane representation from perception, and then holds this command for the following two steps. This is because when new representation is not available, the outdated one is still used by the controller which outputs the same command. This is a typical and easy solution when characterizing lanes using function representation \eqref{eqn:poly_func_repr_D} directly, since the evolution of coefficients in such representation is not obtainable as explained in Section~\ref{sec:veh_ctrl_setup}.

In panel (a), the dotted black curve denotes the closed path (\ref{eqn:EOM_ref_path_ds}, \ref{eqn:curv_s_func}), while the solid red curve represents the position of the vehicle at point $Q$. Panel (b) plots the lateral deviation $\varepsilon_{\Omega}$ using the blue curve, and the steering command $\gamma_{\rm des}$ using the red curve with the axis marked on the right. Panels (c, d, e) show the time profiles of the coefficients $\hat{\phi}_{0}$, $\hat{\phi}_{1}$ and $\hat{\phi}_{2}$, respectively, of the parametric representation \eqref{eqn:param_repr_body_frame} when the vehicle moves along the path. The blue curves denote the measurement outputs of perception algorithm, while the red curves indicate the true values of those coefficients. We remark that: i) only the coefficients until the second order in the lateral direction are shown here since they contain more important information in lateral control than other coefficients; ii) although this represents the scenario using function representation, the coefficients are still transformed to facilitate comparison against parametric representation; and iii) the time profiles of other coefficients in the representation \eqref{eqn:param_repr_body_frame} are similar and reveal similar results.
Fig.~\ref{fig:sim_50ms}(a,b) indicate that the controller (\ref{eqn:lateral_controller}-\ref{eqn:theta0_des}) allows the vehicle to follow the given path with reasonable performance in this scenario. However, during simulation we also observe that when $T_{\rm p}\ge 0.2$ [s], vehicle using this controller is not able to follow the given path using the outdated lane information without prediction. This result implies a high requirement on lane perception latency especially when measurements may also be corrupted with noises in practice.

\begin{figure}
  \centering
  \includegraphics[scale=1.05]{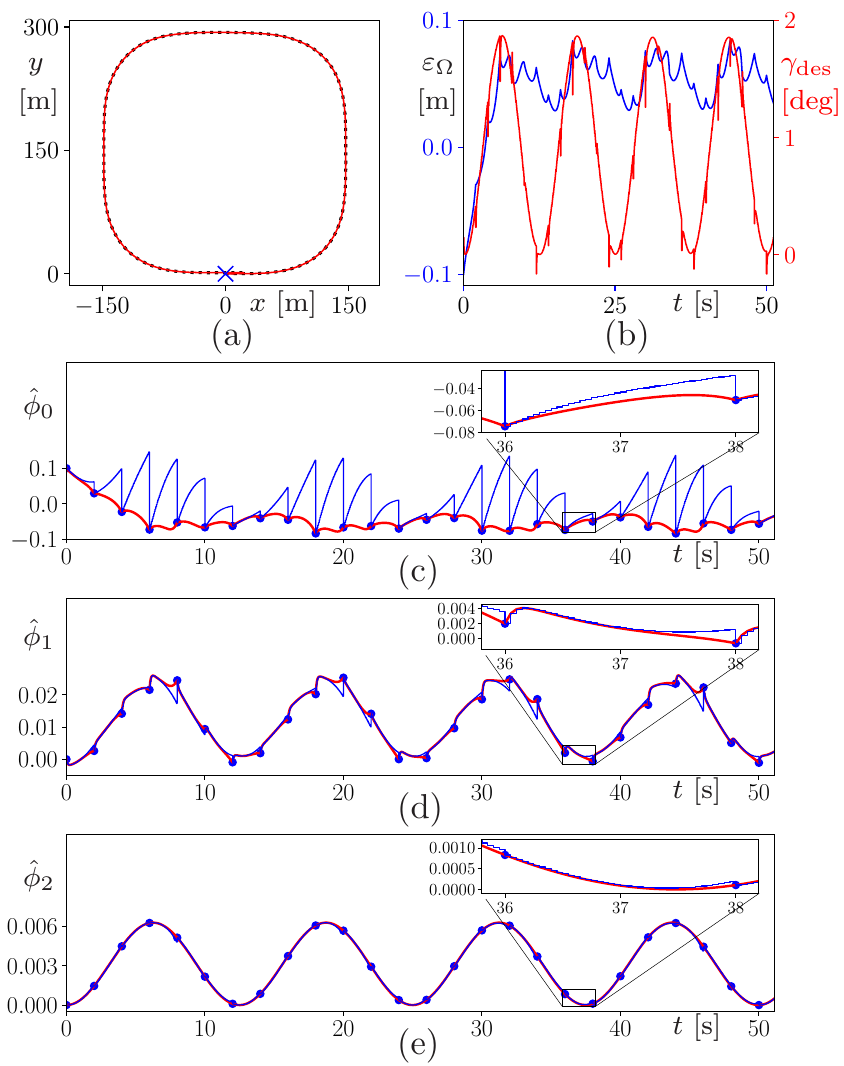}\\
  \caption{Simulation results for camera-based lateral control using \eqref{eqn:coeff_dyn_gen_equiv} for estimation on lane representation when $T=0.05$ [s] and $T_{\rm p}=2$ [s]. (a) Vehicle position and the closed path (\ref{eqn:EOM_ref_path_ds}, \ref{eqn:curv_s_func}) in ($x$, $y$)-plane. (b) lateral deviation $\varepsilon_{\Omega}$ and steering command $\gamma_{\rm des}$. (c, d, e) time profiles of coefficients in the parametric representation \eqref{eqn:param_repr_body_frame} while the vehicle moves along the path. \label{fig:sim_2s}}
\end{figure}

Fig.~\ref{fig:sim_2s} shows the simulation results when function representation \eqref{eqn:poly_func_repr_D} is transformed to parametric representation and \eqref{eqn:coeff_dyn_gen_equiv} is used to predict the coefficients at every control step in the absence of measurement. Here, the control period $T$ is still $0.05$ [s], but perception period $T_{\rm p}$ is set to $2$ [s] to demonstrate the effectiveness. Besides the same layout and color scheme as those used in Fig.~\ref{fig:sim_50ms}, panels (c,d,e) use blue dots to represent new measurements of coefficients from perception algorithm, and also zoomed-in plots to highlight the errors on coefficients between the predicted values and the true values. Fig.~\ref{fig:sim_2s}(a,b) indicate that with such low perception updating rate, the controller still achieves reasonable performance in following the given path by predicting the new representation using \eqref{eqn:coeff_dyn_gen_equiv}. Fig.~\ref{fig:sim_2s}(c,d,e) show that prediction errors on the coefficients of higher orders are much smaller than those of lower orders. This is natural because errors on higher orders lead to less accurate prediction for the future than errors on lower orders. The zoomed-in plots illustrate that the estimation errors in the coefficients  $\hat{\phi}_{0}$, $\hat{\phi}_{1}$ and $\hat{\phi}_{2}$ become noticeable after predicting with \eqref{eqn:coeff_dyn_gen_equiv} for about 0.3, 1.2, and 1.6 [s], respectively. The error in $\hat{\phi}_{0}$ leads to a noticeable offset in lateral deviation $\varepsilon_{\Omega}$ around $0.05$ [m] depicted in Fig.~\ref{fig:sim_2s}(b).
One may notice that the vehicle travels 40 [m] during $2$ [s] prediction period since $V=20$ [m/s]. Function representation \eqref{eqn:poly_func_repr_D} is the Taylor approximation about point $\Omega$ at the perception updating instant. The approximation error gradually becomes noticeable while the vehicle moves forward. Also, prediction with \eqref{eqn:coeff_dyn_gen_equiv} requires estimated state changes (\ref{eqn:changes_tau_eta_psi}, \ref{eqn:changes_s}) using Euler integration that aggregates errors. Although the errors in Taylor approximation cannot be mitigated, the aggregation errors in Euler integration can be reduced by increasing the estimation frequency.

\begin{figure}
  \centering
  \includegraphics[scale=1.05]{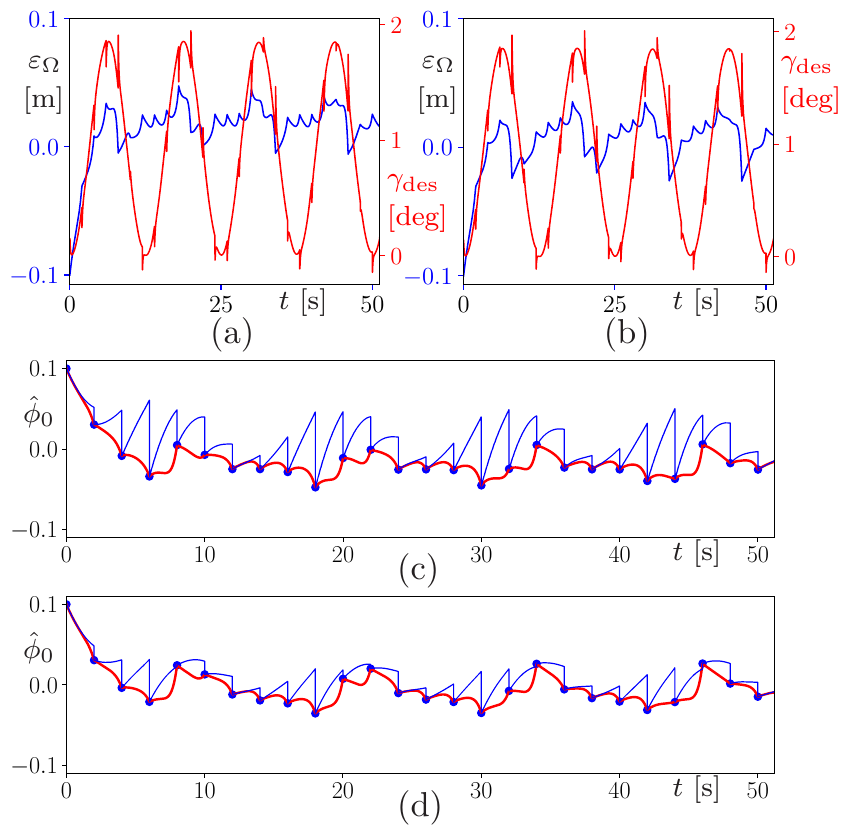}\\
  \caption{Comparison results for camera-based lateral control using \eqref{eqn:coeff_dyn_gen_equiv} for estimation on lane representation when $T_{\rm p}=2$ [s]. (a, c) $T=0.02$ [s]. (b, d) $T=0.01$ [s]. \label{fig:sim_T_diff}}
\end{figure}

Fig.~\ref{fig:sim_T_diff} shows the comparison results for the same scenario as that for Fig.~\ref{fig:sim_2s} when perception updating period $T_{\rm p}$ is still $2$ [s], but control updating period $T$ is reduced to $0.02$ [s] and $0.01$ [s] for panels (a,c) and (b,d), respectively.
Fig.~\ref{fig:sim_2s}(c) and Fig.~\ref{fig:sim_T_diff}(c,d) show that the error of $\hat{\phi}_{0}$ is decreased as control updating period $T$ decreases, which leads to decreased offset in lateral deviation $\varepsilon_{\Omega}$ depicted in Fig.~\ref{fig:sim_2s}(b) and Fig.~\ref{fig:sim_T_diff}(a,b).

We remark that in practice perception updating period is faster than $2$ [s], but the simulations shown here demonstrate the large potential in estimating lanes using parametric representation with \eqref{eqn:coeff_dyn_gen_equiv}. This prediction can be used in the time update step in Kalman filter to get more accurate estimations, or as a pure prediction when measurements are temporarily not available. When noises and model mismatches appear in practice, performance degradation is inevitable. However, it is expected that prediction within $0.5\sim 1$ [s] can still provide reasonable performance since integration aggregation is not long and the vehicle travelling distance is not far.

\section{Conclusion \label{sec:conclusion}}

This paper revisited the fundamental mathematics on approximating curves as polynomial functions or parametric curves. It is shown that arc-length-based parametric representations possess the nice properties of preserving the form in coordinate transformation and parameter shifting. These properties have the potential in facilitating lane estimation for vehicle control since lanes are characterized as curves expressed in vehicle body-fixed frame by perception algorithms. As the vehicle moves, the body-fixed frame is translating and rotating. Thus, we proposed a new architecture using parametric representation in lane estimation and control. To ensure compatibility with most of current platforms, perception algorithms are still assumed to output coefficients of polynomial function representation. We derived the change of coefficients to transform polynomial function representation to arc-length-based parametric representation, and the evolution of coefficients using parametric representation. This evolution reveals an intrinsic linear relationship as the vehicles moves, which can be easily used for prediction or integrated with Kalman filters. We also set up a framework to simulate the whole process, including perception, estimation and control, for camera-based vehicle control problems. Simulation results indicate that controllers relying on predicted lanes using parametric representations can still achieve reasonably good performance at extremely low perception updating rate. These results are practically important in improving control performance with reduced perception updating rate, and obtaining better estimates when coefficients of representations are corrupted with noises. Future research directions may include lane estimation in the presence of noises and model mismatches, and field implementation with the proposed architecture and estimation model.


\bibliographystyle{IEEEtran}

\appendix

\subsection{Coordinates Transformation \label{append:coord_transf}}
\begin{figure}
  \centering
  \includegraphics[scale=0.5]{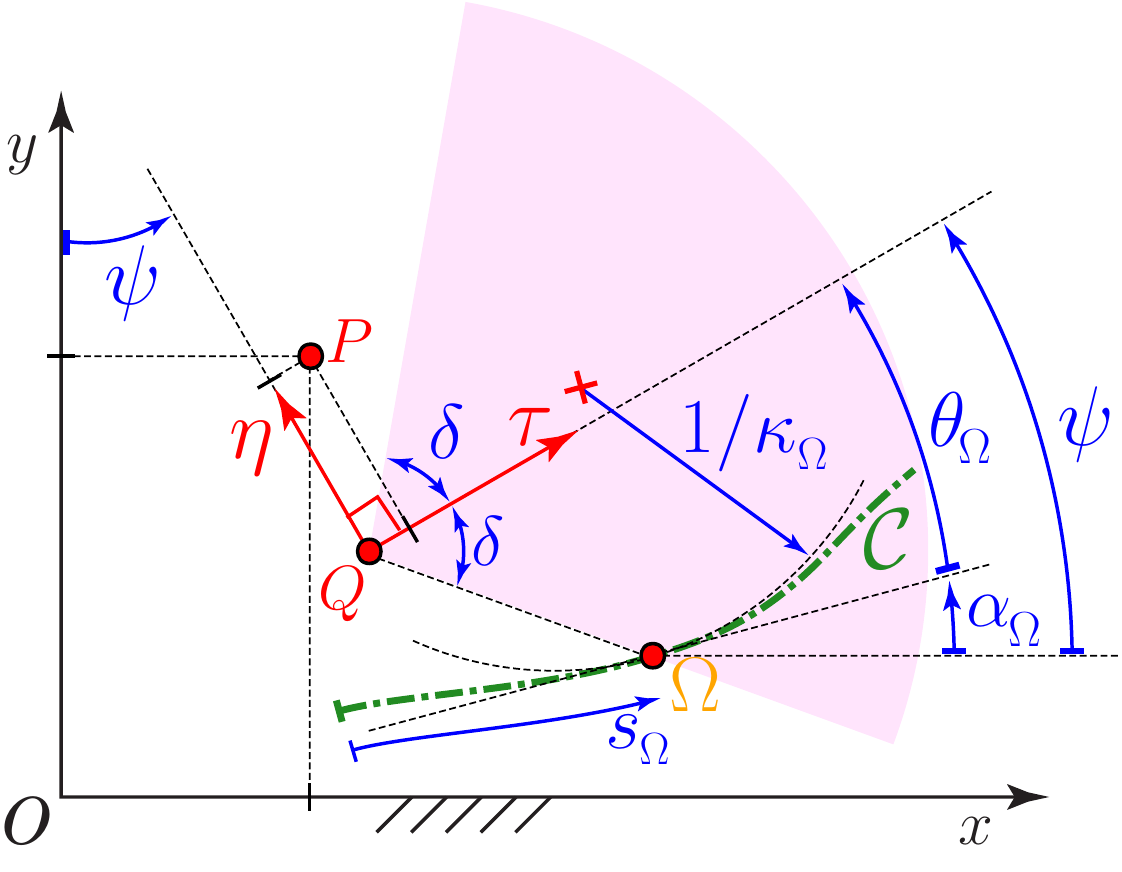}\\
  \caption{coordinates transformation.}\label{fig:coord_transf}
\end{figure}

Let us consider an arbitrary point $P$ (cf.~Fig.~\ref{fig:coord_transf}) whose position are given by ${(x,y)}$ and ${(\tau, \eta)}$ in $\cFE$ and $\cFB$, respectively. Applying Theorem~\ref{thm:gen_coord_transf} yields the change of coordinates
\begin{subequations}\label{eqn:transf_xy_2_taueta}
\begin{align}
 \tau &= (x-x_{Q})\cos\psi+(y-y_{Q})\sin \psi\, ,\label{eqn:transf_xy_2_tau}\\
 \eta & = -(x-x_{Q})\sin\psi+(y-y_{Q})\cos \psi\, .\label{eqn:transf_xy_2_eta}
\end{align}
\end{subequations}
The definition of state $\varepsilon_{\Omega}$ leads to
\begin{equation}\label{eqn:lat_dev_def}
  \begin{split}
    \pvec{\Omega Q}\cdot \unitvec[\tau] &= -|\varepsilon_{\Omega}|\cos\delta\,,\qquad
  \pvec{\Omega Q}\cdot \unitvec[\eta]=\varepsilon_{\Omega}\sin \delta\,,
  \end{split}
\end{equation}
implying that
\begin{subequations}
\begin{align}
  (x_{Q}-x_{\Omega})\cos\psi+(y_{Q}-y_{\Omega})\sin\psi &= -|\varepsilon_{\Omega}|\cos\delta\,,\label{eqn:orgin_colinear_cond}\\
  -(x_{Q}-x_{\Omega})\sin\psi+(y_{Q}-y_{\Omega})\cos\psi&=\varepsilon_{\Omega}\sin \delta\,.\label{eqn:lat_dev_exp}
\end{align}
\end{subequations}
Adding (\ref{eqn:orgin_colinear_cond}, \ref{eqn:lat_dev_exp}) to (\ref{eqn:transf_xy_2_tau}, \ref{eqn:transf_xy_2_eta}), respectively, we obtain
\begin{equation}\label{eqn:transf_xy_2_tau_eta_new}
  \begin{split}
  \tau &= (x-x_{\Omega})\cos\psi+(y-y_{\Omega})\sin \psi+|\varepsilon_{\Omega}|\cos\delta\, ,\\
 \eta & = -(x-x_{\Omega})\sin\psi+(y-y_{\Omega})\cos \psi-\varepsilon_{\Omega}\sin\delta\, ,
  \end{split}
\end{equation}
and their corresponding derivatives are
\begin{equation}\label{eqn:transf_tau_eta_deriv}
  \begin{split}
    \dot{\tau} & =(\dot{x}-\dot{x}_{\Omega})\cos\psi+(\dot{y}-\dot{y}_{\Omega})\sin \psi\\
  &+(\eta+\varepsilon_{\Omega}\sin \delta)\dot{\psi}+\dot{\varepsilon}_{\Omega}\,\sign(\varepsilon_{\Omega})\cos\delta\, ,
 \\
 \dot{\eta} & =-(\dot{x}-\dot{x}_{\Omega})\sin\psi+(\dot{y}-\dot{y}_{\Omega})\cos \psi\\
 &-(\tau-|\varepsilon_{\Omega}|\cos\delta)\dot{\psi}-\dot{\varepsilon}_{\Omega}\sin\delta\, .
  \end{split}
\end{equation}

Notice that $\Omega$ is a point on the path and
\begin{equation}\label{eqn:EOM_ref_path_dt}
\begin{split}
    \dot{x}_{\Omega} &= \dfrac{\textrm{d} x_{\Omega}}{\textrm{d} t} =\dfrac{\textrm{d} x_{\Omega}}{\textrm{d} s_{\Omega}}\dfrac{\textrm{d} s_{\Omega}}{\textrm{d} t} = \cos\alpha_{\Omega}\dot{s}_{\Omega}\, ,
    \\
    \dot{y}_{\Omega} &= \dfrac{\textrm{d} y_{\Omega}}{\textrm{d} t} =\dfrac{\textrm{d} y_{\Omega}}{\textrm{d} s_{\Omega}}\dfrac{\textrm{d} s_{\Omega}}{\textrm{d} t} = \sin\alpha_{\Omega}\dot{s}_{\Omega}\, ,
    \\
    \dot{\alpha}_{\Omega} &=\dfrac{\textrm{d} \alpha_{\Omega}}{\textrm{d} t} =\dfrac{\textrm{d} \alpha_{\Omega}}{\textrm{d} s_{\Omega}}\dfrac{\textrm{d} s_{\Omega}}{\textrm{d} t} =\kappa_{\Omega}\dot{s}_{\Omega}\, .
\end{split}
\end{equation}
Substituting \eqref{eqn:EOM_ref_path_dt} into \eqref{eqn:transf_tau_eta_deriv} and combining the derivative of \eqref{eqn:rel_heading_def} yield the relationship
\begin{equation}\label{eqn:transf_s_eta_deriv_2}
\begin{split}
 \dot{\tau} & =\dot{x}\cos\psi + \dot{y}\sin \psi -\dot{s}_{\Omega}\cos\theta_{\Omega}\\
 &+(\eta+\varepsilon_{\Omega}\sin \delta)\dot{\psi}+\dot{\varepsilon}_{\Omega}\,\sign(\varepsilon_{\Omega})\cos\delta\, ,
 \\
 \dot{\eta} & =-\dot{x}\sin\psi+\dot{y}\cos \psi+\dot{s}_{\Omega}\sin\theta_{\Omega}\\
 &-(\tau-|\varepsilon_{\Omega}|\cos\delta)\dot{\psi}-\dot{\varepsilon}_{\Omega}\sin\delta\, ,\\
 \dot{\theta}_{\Omega} &= \dot{\psi}-\kappa_{\Omega}\dot{s}_{\Omega}\,,
\end{split}
\end{equation}

Solving  \eqref{eqn:transf_s_eta_deriv_2}, we obtain the transformation of a general point from absolute position ($x$ and $y$) and orientation ($\psi$) to relative position ($s_{\Omega}$ and $\varepsilon_{\Omega}$) and orientation $\theta_{\Omega}$ with respect to a path as
\begin{equation}
  \begin{split}\label{eqn:transf_xy_se_diff_gen}
    \dot{s}_{\Omega} &=\dfrac{\sin(\delta-\psi)}{\sin(\delta-\theta_{\Omega})} \dot{x}+ \dfrac{\cos(\delta-\psi)}{\sin(\delta-\theta_{\Omega})}\dot{y}\\
    &+\dfrac{|\varepsilon_{\Omega}|\cos\big(\delta-\sign(\varepsilon_{\Omega})\,\delta\big)-\tau\cos\delta+\eta \sin\delta}{\sin(\delta-\theta_{\Omega})} \dot{\psi}\\
    &-\dfrac{\sin\delta}{\sin(\delta-\theta_{\Omega})} \dot{\tau}-\dfrac{\cos\delta}{\sin(\delta-\theta_{\Omega})} \dot{\eta}\,,\\
    \dot{\varepsilon}_{\Omega} &=-\dfrac{\sin(\psi-\theta_{\Omega}) }{\sin(\delta-\theta_{\Omega})}\dot{x}+ \dfrac{\cos(\psi-\theta_{\Omega})}{\sin(\delta-\theta_{\Omega})}\dot{y}\\
    &+\dfrac{|\varepsilon_{\Omega}| \cos\big(\delta-\sign(\varepsilon_{\Omega})\,\theta_{\Omega}\big)-\tau \cos\theta_{\Omega}+\eta\sin\theta_{\Omega}}{\sin(\delta-\theta_{\Omega})}\dot{\psi}\\
    &-\dfrac{\sin\theta_{\Omega}}{\sin(\delta-\theta_{\Omega})}\dot{\tau}-\dfrac{\cos\theta_{\Omega}}{\sin(\delta-\theta_{\Omega})}\dot{\eta}\,,\\
    \dot{\theta}_{\Omega} &=-\dfrac{\kappa_{\Omega} \sin(\delta-\psi)}{\sin(\delta-\theta_{\Omega})} \dot{x}-\dfrac{\kappa_{\Omega} \cos(\delta-\psi)}{\sin(\delta-\theta_{\Omega})}\dot{y}\\
    &+\Big(1-\dfrac{\kappa_{\Omega}\,|\varepsilon_{\Omega}|\cos\big(\delta-\sign(\varepsilon_{\Omega})\,\delta\big)}{\sin(\delta-\theta_{\Omega})}\\
    &+\dfrac{\kappa_{\Omega}(\tau \cos\delta-\eta \sin\delta)}{\sin(\delta-\theta_{\Omega})}\Big) \dot{\psi}
    +\dfrac{\kappa_{\Omega} \sin\delta}{\sin(\delta-\theta_{\Omega})}\dot{\tau}\\
    &+\dfrac{\kappa_{\Omega} \cos\delta}{\sin(\delta-\theta_{\Omega})}\dot{\eta}\,,
  \end{split}
\end{equation}
in the differential format, and its inverse transformation is
\begin{equation}
  \begin{split}\label{eqn:transf_se_xy_diff_gen}
    \dot{x} &= \Big(\cos\alpha_{\Omega}-\kappa_{\Omega}\varepsilon_{\Omega} \sin\big(\delta-\sign(\varepsilon_{\Omega})\, \psi\big)-\kappa_{\Omega}\,\tau\sin\psi\\
    &-\kappa_{\Omega}\,\eta  \cos\psi\Big) \dot{s}_{\Omega}-\big(\varepsilon_{\Omega} \sin\big(\delta-\sign(\varepsilon_{\Omega})\,\psi\big)+\tau\sin\psi\\
    &+\eta \cos\psi\big) \dot{\theta}_{\Omega}-\dot{\varepsilon}_{\Omega} \cos(\delta-\psi)+\dot{\tau} \cos\psi-\dot{\eta}\sin\psi\,,\\
    \dot{y} &= \Big(\sin\alpha_{\Omega}- \kappa_{\Omega}\, |\varepsilon_{\Omega}| \cos\big(\delta-\sign(\varepsilon_{\Omega})\, \psi\big)+\kappa_{\Omega}\,\tau \cos\psi\\
    &-\kappa_{\Omega}\,\eta\sin\psi\Big) \dot{s}_{\Omega}-\big(|\varepsilon_{\Omega}| \cos\big(\delta-\sign(\varepsilon_{\Omega})\, \psi\big)-\tau \cos\psi\\
    &+\eta\sin\psi\big) \dot{\theta}_{\Omega}+\dot{\varepsilon}_{\Omega} \sin(\delta-\psi)+\dot{\tau}\sin\psi +\dot{\eta} \cos\psi\,,\\
    \dot{\psi} &= \kappa_{\Omega} \dot{s}_{\Omega}+\dot{\theta}_{\Omega}\,.
  \end{split}
\end{equation}
Solving (\ref{eqn:rel_heading_def},\ref{eqn:transf_xy_2_tau_eta_new}) simultaneously, one can also obtain the inverse transformation
\begin{equation}\label{eqn:transf_se_xy_gen}
  \begin{split}
    x &= x_{\Omega}-|\varepsilon_{\Omega}| \cos\big(\alpha_{\Omega}+\theta_{\Omega}-\sign(\varepsilon_{\Omega})\,\delta\big)\\
    &+\tau \cos(\alpha_{\Omega}+\theta_{\Omega})-\eta\sin(\alpha_{\Omega}+\theta_{\Omega})\,, \\
    y &= y_{\Omega}-|\varepsilon_{\Omega}| \sin\big(\alpha_{\Omega}+\theta_{\Omega}-\sign(\varepsilon_{\Omega})\,\delta\big)\\
    &+\tau\sin(\alpha_{\Omega}+\theta_{\Omega})+\eta \cos(\alpha_{\Omega}+\theta_{\Omega})\,,\\
    \psi &= \alpha_{\Omega}+\theta_{\Omega}\,,
  \end{split}
\end{equation}
expressed in coordinates form, where $x_{\Omega}$, $y_{\Omega}$ and $\alpha_{\Omega}$ depend on $s_{\Omega}$. We remark that it is hard to explicitly express the transformation \eqref{eqn:transf_xy_se_diff_gen} in the coordinates form.

Setting the location $x=x_{\rm Q}$,  $y=y_{\rm Q}$, $\tau\equiv 0$ and $\eta \equiv 0$ and utilizing the equation of motion of the camera location $Q$ given in \eqref{eqn:EOM_point_Q},
we obtain the relative equation of motion \eqref{eqn:transf_xy_se_diff_Q} with respect to the path and the inverse transformation is
\begin{equation}\label{eqn:transf_se_xy_Q}
  \begin{split}
    x_{\rm Q} &= x_{\Omega}-|\varepsilon_{\Omega}|\cos\big(\alpha_{\Omega}+\theta_{\Omega}-\sign(\varepsilon_{\Omega})\,\delta\big)\,,\\
    y_{\rm Q} &= y_{\Omega}-|\varepsilon_{\Omega}|\sin\big(\alpha_{\Omega}+\theta_{\Omega}-\sign(\varepsilon_{\Omega})\,\delta\big)\,,\\
    \psi &= \alpha_{\Omega}+\theta_{\Omega}\,.
  \end{split}
\end{equation}

\subsection{Proof of Lemma~\ref{lemma:gen_curve_func_repr} \label{append:path_func_repr_body_frame}}
One can apply Theorem~\ref{thm:gen_curve_param_repr} to obtain the parametric representation about point $\Omega$ in frame $\cFE$, and then apply Theorem~\ref{thm:curve_param_repr_conformal} to transform this representation to frame $\cFB$. Finally the function representation in body-fixed frame can be obtained by applying Theorem~\ref{thm:curve_param_2_func_repr}. Alternatively, the function representation in body-fixed frame can also be calculated directly as follows.

Given an arbitrary point on the curve whose coordinates are $(x, y)$ and $(\tau, \eta)$ in frame $\cFE$ and $\cFB$, respectively, applying Theorem~\ref{thm:gen_coord_transf} leads to the same coordinate transformation as that given in \eqref{eqn:transf_xy_2_taueta}.
Notice that vehicle position $(x_{Q}, y_{Q})$ and orientation $\psi$ are independent of arc-length parameter $s$, while the relationship between $(x, y)$ and $s$ are given in \eqref{eqn:EOM_ref_path_ds} based on differential geometry. Thus, differentiating (\ref{eqn:transf_xy_2_taueta}) with respect to $s$ yields
\begin{equation}\label{eqn:eta_tau_derivs}
  \begin{split}
    \tau' &= \cos(\alpha-\psi)\,,\quad\hspace{46pt}
    \eta' =\sin(\alpha-\psi)\,,\\
    \tau''&=-\kappa\sin(\alpha-\psi)\,,\qquad\qquad
    \eta''=\kappa\cos(\alpha-\psi)\,,\\
    \tau'''&=-\kappa^{2}\cos(\alpha-\psi)-\kappa'\sin(\alpha-\psi)\,,\\
    \tau^{(4)}&=-(\kappa''-\kappa^{3})\sin(\alpha-\psi)-3\kappa\kappa'\cos(\alpha-\psi)\,,\\
    \tau^{(5)}&=\big(\kappa^{4}-3(\kappa')^{2}-4\kappa\kappa''\big)\cos(\alpha-\psi)\\
    &+(6\kappa^{2}\kappa'-\kappa''')\sin(\alpha-\psi)\\
    \eta''' &=-\kappa^{2}\sin(\alpha-\psi)+\kappa'\cos(\alpha-\psi)\,,\\
    \eta^{(4)}&=(\kappa''-\kappa^{3})\cos(\alpha-\psi)-3\kappa\kappa'\sin(\alpha-\psi)\,,\\
    \eta^{(5)}&=\big(\kappa^{4}-3(\kappa')^{2}-4\kappa\kappa''\big)\sin(\alpha-\psi)\\
    &-(6\kappa^{2}\kappa'-\kappa''')\cos(\alpha-\psi)\,.
  \end{split}
\end{equation}
Also, notice that
\begin{equation}\label{eqn:proof_deta_dtau}
    \dfrac{\diff \eta}{\diff \tau} =\dfrac{\eta'}{\tau'}\,,
\end{equation}
and
\begin{equation}\label{eqn:proof_deta_dtau_n}
    \dfrac{\diff^{n+1} \eta}{\diff \tau^{n+1}} =\dfrac{\diff}{\diff \tau}\left(\dfrac{\diff^{n} \eta}{\diff \tau^{n}}\right)
    =\dfrac{\left(\frac{\diff^{n} \eta}{\diff \tau^{n}}\right)'}{ \tau'}\,,\; n=1,\, 2,\, 3,\, \ldots\,.
\end{equation}
Substituting \eqref{eqn:eta_tau_derivs} into (\ref{eqn:proof_deta_dtau}, \ref{eqn:proof_deta_dtau_n}), one can obtain the derivatives recursively until the required order. Then the coefficients of function representation can be obtained by evaluating these derivatives at point $\Omega$, and utilizing (\ref{eqn:rel_heading_def}, \ref{eqn:func_repr_coeff_def}).


\begin{IEEEbiography}
[{\includegraphics[width=1in,height=1.25in,clip,keepaspectratio]{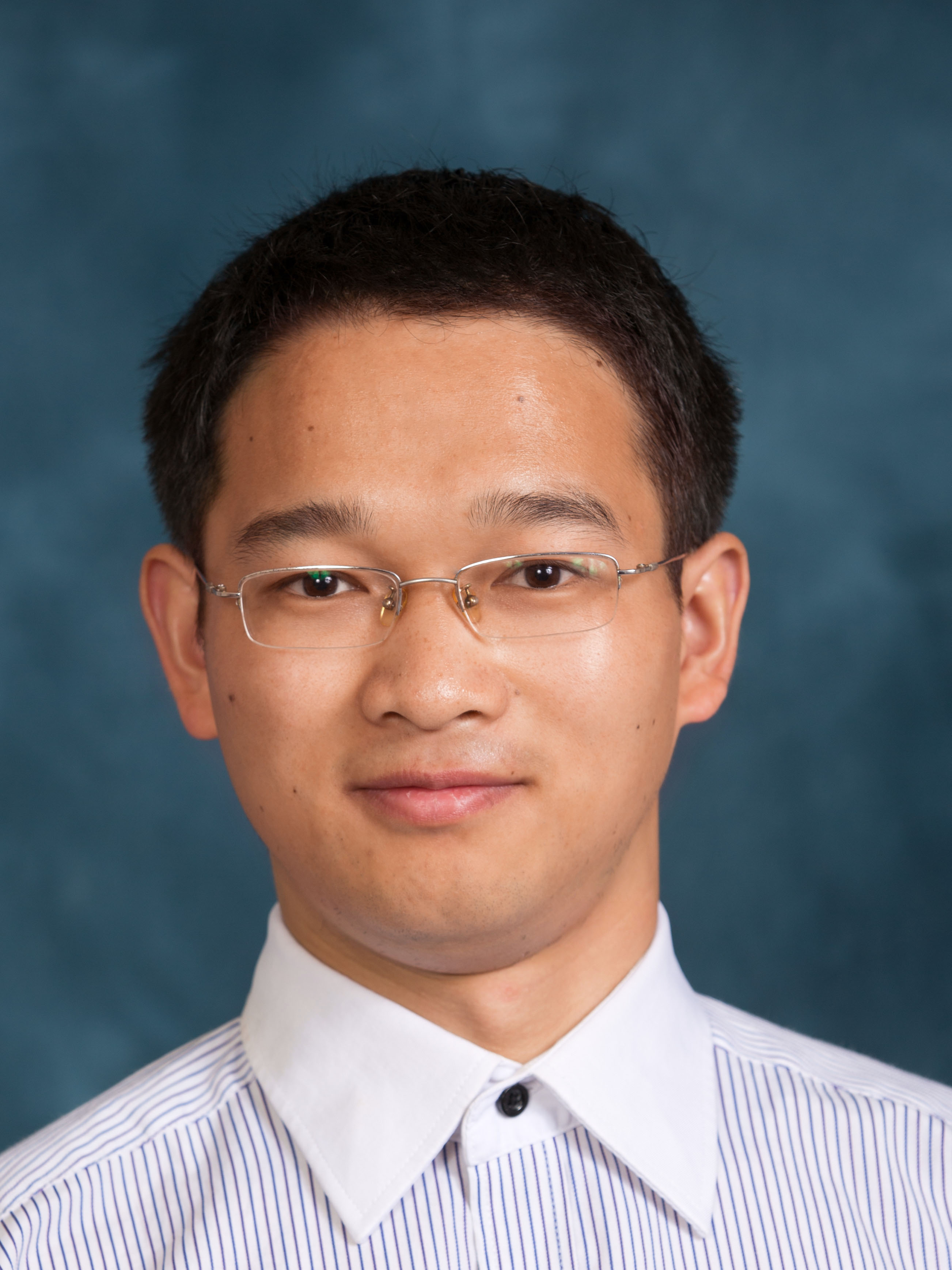}}]{Wubing B. Qin}
received his BEng degree in School of Mechanical Science and Engineering from Huazhong University of Science and Technology, China in 2011, and his MSc degree and PhD degree in Mechanical Engineering from the University of Michigan, Ann Arbor in 2016 and 2018, respectively. Curretly he is an independent researcher without affiliations. His research focuses on dynamics, control theory, connected/automated vehicles, ground robotics, mechatronics and nonlinear systems.
\end{IEEEbiography}

\end{document}